\documentclass[twocolumn,showpacs,preprintnumbers,amsmath,amssymb
            ,prb,eqsecnum]{revtex4}

\usepackage{graphicx}
\usepackage{dcolumn}
\usepackage{bm}

\begin{document}

\preprint{l-edge-magnon}

\title{Magnetic excitations in L-edge resonant inelastic\\
x-ray scattering from cuprate compounds}

\author{Jun-ichi Igarashi}%
\affiliation{%
Faculty of Science, Ibaraki University, Mito, Ibaraki 310-8512, Japan}

\author{Tatsuya Nagao}
\affiliation{%
Faculty of Engineering, Gunma University, Kiryu, Gunma 376-8515, Japan}

\date{\today}

\begin{abstract}

We study the magnetic excitation spectra in 
$L$-edge resonant inelastic
x-ray scattering (RIXS) from undoped cuprates.
We analyze the second-order dipole allowed process that 
the strong perturbation works through the intermediate state
in which the spin degree of freedom is lost at the core-hole site. 
Within the approximation neglecting the perturbation on the neighboring sites, 
we derive the spin-flip final state in the scattering channel with changing
the polarization, which leads to the RIXS 
spectra expressed as the dynamical structure factor of the transverse 
spin components. We assume a spherical form of the spin-conserving final state
in the channel without changing the polarization, 
which leads to the RIXS spectra expressed as the
'exchange'-type multi-spin correlation function.
Evaluating numerically the transition amplitudes to these final states
on a finite-size cluster, we obtain a sizable amount of the transition 
amplitude to the spin-conserving final state in comparison with that 
to the spin-flip final state.
We treat the itinerant magnetic 
excitations in the final state by means of the $1/S$-expansion method. 
Evaluating the higher-order correction with $1/S$, we find that
the peak arising from the one-magnon excitation is reduced with its weight,
and the continuous spectra arising from the three-magnon excitations come out.
The interaction between two magnons is treated by summing up the ladder
diagrams. On the basis of these results, we analyze the $L_3$-edge RIXS spectra 
in Sr$_2$CuO$_2$Cl$_2$ in comparison with the experiment.
It is shown that the three-magnon excitations as well as the two-magnon
excitations give rise to the intensity in the high energy side of 
the one-magnon peak, making the spectral shape asymmetric with wide width, 
in good agreement with the experiment.

\end{abstract}

\maketitle

\section{\label{sect.1}Introduction}

Resonant inelastic x-ray scattering (RIXS) has recently attracted
much interest as a useful tool to investigate excited states in solids. 
The $K$- and $L$-edge resonances have been widely used in transition-metal 
compounds. 
When one tries to interpret the experimental results, some theoretical support
becomes inevitable.
Since the intermediate states involved in the scattering process
at the $K$- and $L$-edges are different from each other, the theoretical
analysis should discriminate the difference with great care.

The $K$-edge resonance is more suitable than the $L$-edge to detect 
the momentum dependence of the spectra, since the wavelength of the $K$-edge x-ray
is an order of the lattice constant.
In the $K$-edge RIXS, the $1s$-core electron is prompted to an empty $4p$ state
by absorbing photon, then charge excitations are created in order to screen 
the core-hole potential, and finally the photo-excited $4p$ electron is
recombined with the core hole by emitting photon.
Charge excitations are finally left with energy and momentum transferred from photons.
\cite{Kao96,Hill98,Hasan00,Kim02,Inami03,Kim04-1,Suga05} 
The spectra have been analyzed by several methods.
\cite{Tsutsui99,Okada06,vdBrink06,Ament07} 
Among them, a formalism developed by Nomura and Igarashi (NI)
\cite{Nomura04,Nomura05,Igarashi06} usefully describes the RIXS 
spectra in terms of the $3d$-density-density correlation function 
on the initial state.\cite{Com3}
It is based on the Keldysh Green function,\cite{Keldysh65} 
and is regarded as
an adaption of the resonant Raman theory of Nozi\`eres and Abrahams.
\cite{Nozieres74} 
In the actual application of the formalism, the electronic states are described 
within the Hartree-Fock approximation on the multiband tight-binding model, 
and the electron correlation is treated within the random phase approximation.
The spectra are calculated as functions of energy loss and momentum 
transfer in good agreement with the experiments
for undoped cuprates,\cite{Nomura04,Nomura05,Igarashi06}
NiO,\cite{Takahashi07} and LaMnO$_3$.\cite{Semba08}

In addition to the charge excitations, the magnetic excitations have been 
observed from the Cu $K$-edge RIXS experiment in La$_2$CuO$_4$.
\cite{Hill08,Ellis10}
They are brought about through the modification of the exchange coupling 
between the spins of $3d$ electrons at the core-hole site and those 
at the neighboring sites by the core-hole potential.\cite{vdBrink05,vdBrink07,Forte08}
To investigate the magnetic excitation spectrum,
the NI formula has been adapted in our previous work.\cite{Nagao07}
By replacing the channel of 
creating an electron-hole pair by that of creating two magnons, 
the expression is derived that the RIXS spectra are proportional 
to the two-magnon correlation function.
The magnon-magnon interaction in the correlation function 
has been treated with the use of the $1/S$-expansion method
($S$ is the magnitude of spin).
\cite{Oguchi60,Harris71,Hamer92,Canali92,Igarashi92-1,Igarashi92-2}
The obtained spectra, reflecting a significant influence from
the magnon-magnon interaction, show characteristic dependence on
the energy loss and the momentum transfer, which captures features
the experimental data have demonstrated.\cite{Hill08,Ellis10}

\begin{figure}
\includegraphics[width=8.0cm]{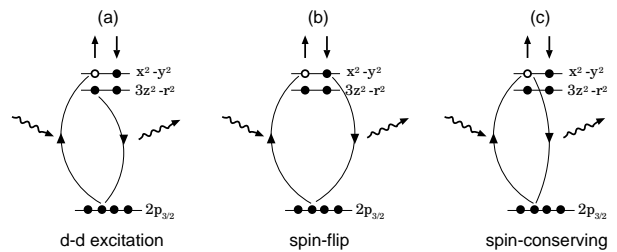}%
\caption{\label{fig.process}
$L_3$-edge RIXS process for undoped cuprates.
Black and white circles represent electrons and holes, respectively.
Vertical arrows indicate the spin direction.
Wavy lines indicate the incident and scattered photon.
The $t_{2g}$ orbitals are omitted.
}
\end{figure}

Unlike the $K$-edge RIXS, the $L$-edge RIXS directly accesses the $3d$ states,
which leads us to a necessity to develop an appropriate treatment for
the $L$-edge RIXS.
The process is illustrated for undoped cuprates in Fig.~\ref{fig.process};
the $2p$-core electron is prompted to the empty
$x^2-y^2$ orbital by absorbing photon, and then an occupied $3d$ electron
combines with the core hole by emitting photon.
Here $x$, $y$, and $z$ axes correspond to the crystal axes $a$, $b$, and $c$, 
respectively.
If the $3d$ state in the photo-emitting process is different from
the $x^2-y^2$ orbital [Fig.~\ref{fig.process} (a)], 
excitations within the $3d$ states are left
in the final state, which process is called the `d-d' transition.
\cite{Ghiringhelli04}
Since such excited states are quite localized at the core-hole site,
the spectra show little momentum dependence.
In this paper, skipping the study of the $d$-$d$ transition, we concentrate 
on the study of the $L$-edge RIXS spectra arising from magnetic excitations, 
which have recently been observed in undoped cuprates.
\cite{Braicovich09,Braicovich10,Guarise10}

It is known that the spin-flip excitations [Fig.~\ref{fig.process}(b)]
give rise to the spin-wave-like dispersion,
when the direction of the staggered moment deviates from the $z$ 
direction.\cite{Ament09} 
Since the spin angular-momentum is coupled to the orbital 
angular-momentum through the large spin-orbit interaction of the $2p$-core 
states, the change by the spin-flip may be compensated by the change of 
the polarization of photon. In addition, it is expected that 
the spin-conserving excitations 
[Fig.~\ref{fig.process}(c)] could be brought about through the intermediate 
state, since the strong perturbation is working by losing the spin degree 
of freedom at the core-hole site in the intermediate state.

For investigating the magnetic excitations,
the RIXS spectra have been analyzed by assuming an extreme condition that
the core-hole life-time is so short that the intermediate state
can not have enough time to relax.\cite{Ament07}
Only the one-magnon excitation is brought about without any two-magnon 
excitations. This is called as the ultra-short core-hole life-time (UCL) 
approximation,\cite{Ament09}
which reminds us of the fast collision approximation in 
resonant elastic x-ray  scattering.\cite{Luo93,Carra94} 
In reality, the spectra observed as a function of energy loss exhibit 
a systematic change with changing momentum transfer;\cite{Guarise10}
the peak position moves according to the spin-wave dispersion curve,
while their shapes exhibit structures indicative of two- or three-magnon 
excitations.
Accordingly the assumed condition is obviously unsatisfied in undoped cuprates.
The purpose of this paper is to develop a comprehensive theory describing
the spectra beyond the UCL approximation.

The situation that the spin degree of freedom is lost at the core-hole site
in the intermediate state is similar to the situation of the non-magnetic 
impurity,
which problem has been extensively studied by the linear-spin-wave 
(LSW) method.
\cite{Tonegawa68,Wan93}
The method developed there, however, is not applicable to the present 
problem, since the calculation of the RIXS spectra require 
not alone the intermediate state but also the initial and final
states in evaluating overlaps between them. 
It may not be logically appropriate to perform a perturbation calculation
with the terms involving the lost spin degree of freedom, although
such attempts have been mentioned and tried.\cite{Ament11}
Along this consideration,
the perturbative expansion like the NI formula may not be sufficient 
to treat the intermediate state. 

In this paper, we analyze the second-order dipole allowed process
that the strong perturbation works through the intermediate state.
Within the approximation that the perturbation is not extending to neighboring
sites, we derive the spin-flip final state expressed as
$\mbox{\boldmath{$\alpha$}}_{f\perp}\times
\mbox{\boldmath{$\alpha$}}_{i\perp}\cdot{\bf S}_0|g\rangle$
in the scattering channel with changing the polarization,
where $\mbox{\boldmath{$\alpha$}}_{i (f)\perp}$ is the polarization 
vector of incident (scattered) photon
projected onto the $x$-$y$ plane, and ${\bf S}_0$ is the spin operator vector
at the core-hole site. This form is inferred from the formula of the elastic
scattering.\cite{Hannon1988,Haverkort2010} 
This final state leads to the RIXS spectra expressed as 
the dynamical structure factor of the transverse spin component.
In the scattering channel with changing the polarization,
we assume a spherical form of the spin-conserving final state,
which leads to the RIXS spectra expressed as the 'exchange'-type
multi-spin correlation function.
In the evaluation of the transition amplitude to these final states,
we introduce a finite-size cluster and use the wave-functions 
numerically obtained by diagonalizing the Hamiltonian. 
We obtain a sizable amount of the transition amplitude to the
spin-conserving final state in comparison with that to the spin-flip
final state.

We express the spin-flip and spin-conserving final states in terms of
magnon creations and annihilations with the systematic use of the $1/S$
expansion.
The spin-flip operators $S^{\pm}$ are expanded in terms of one-magnon creation
and annihilation operators within the LSW theory, and of
three-magnon creation and annihilation operators in the second 
order of $1/S$.\cite{Igarashi92-1,Igarashi93,Canali92,Hamer92,Igarashi05}
Hence the spin-flip final state gives rise to the RIXS spectra consist of two
ingredients. One is
the $\delta$-function peak with its weight reduced from the LSW value and
the other is the broad three-magnon continuum as a function of energy loss.
The spin-conserving operators are expanded in terms of two-magnon creation and
annihilation operators with neglecting the higher-order terms
such as four-magnon creation and annihilation operators.
Since two magnons are created near the core-hole site, we take account of their
interaction by summing up the ladder diagrams, as was done in the magnetic 
excitation spectra in the $K$-edge RIXS.\cite{Nagao07}

On the basis of these results, we analyze the Cu $L_3$-edge spectra 
of Sr$_2$CuCl$_2$O$_2$, which show strong dependence on the polarization
in accordance with the experiment.\cite{Guarise10}
We find that a considerable amount of satellite intensity comes from
the three-magnon excitations in the spin-flip final state,
which leads to the asymmetric shape as a function of energy loss
in the $\pi$ polarization.
We also obtain the substantial contribution from the two-magnon excitations
in the $\sigma$ polarization. These results agree well with the experiment.
 
The present paper is organized as follows.
In Sec. \ref{sect.2}, we describe the Hamiltonian responsible for magnetic
excitations and transition-matrix elements relevant to the $L_{2,3}$ edges
in cuprate compounds.
In Sec. \ref{sect.3}, we discuss the process giving rise to the magnetic
excitations through the intermediate state.
The RIXS spectra are expressed in terms of spin-correlation functions.
In Sec. \ref{sect.4}, spin-correlation functions are calculated by
means of the $1/S$ expansion.
In Sec. \ref{sect.5}, the calculated $L_3$-edge spectra of Cu 
are compared with those for Sr$_2$CuO$_2$Cl$_2$.
Section \ref{sect.6} is devoted to the concluding remarks.
In Appendix, a brief summary of the $1/S$-expansion method is given.

\section{\label{sect.2}Formulation of RIXS spectra at the $L_{2,3}$ edge}

\subsection{Hamiltonian}
At the half-filling in cuprate compounds, each Cu atom has 
one hole in the $x^2-y^2$ orbital, where the $x$ and $y$ axes are defined
along the Cu-O bonds and the $z$ along the crystal $c$ axis.
Adopting the hole picture, 
we assume a single band Hubbard model on a 
two-dimensional square lattice
for $3d$ electrons:
\begin{equation}
 H = t\sum_{\langle i,j\rangle, \sigma} (d^{\dagger}_{i\sigma}d_{j\sigma} + {\rm H.c.})
   + U\sum_i d_{i\uparrow}^{\dagger}d_{i\downarrow}^{\dagger}
             d_{i\downarrow}d_{i\uparrow},
\label{eq.Hubbard}
\end{equation}
where $d_{i\sigma}$ ($d_{i\sigma}^{\dagger}$)represents 
the annihilation (creation) operator of the hole
with spin $\sigma$ at site $i$. The sum over 
$\langle i,j\rangle$ extends
over distinct pairs of nearest neighbors. Hopping integral and on-site Coulomb interaction are denoted as $t$ and $U$, respectively.
We have neglected the small inter-layer coupling.  
This Hubbard model may be mapped from a more precise ``d-p" model
for cuprate compounds.

For low-energy spin excitations, the Hubbard Hamiltonian is further mapped onto
the two-dimensional Heisenberg Hamiltonian with 
the exchange coupling constant $J=4t^2/U$,
\begin{equation}
 H_{\rm mag}=J \sum_{\langle i,j\rangle} 
                   \textbf{S}_i\cdot \textbf{S}_{j}.
\end{equation}
This Hamiltonian leads to an antiferromagnetic order. It is known that
Cu spins align on the CuO$_2$ plane at the angle of $\pi/4$ to the 
Cu-O bonds.\cite{Vaknin87}
Therefore, it is convenient to define 
the spin coordinate frame $x'$, $y'$, $z'$ axes by rotating
the crystal-fixed coordinate frame $a$, $b$, $c$ axes
through the Euler angles $\alpha=\pi/4$, $\beta=\pi/2$, and $\gamma=0$.
The $\uparrow$ and $\downarrow$ states in Eq.~(\ref{eq.Hubbard})
are interpreted as the eigenstates with respect to $S^{z'}$. 
Hereafter, we shall describe the spin state of the $3d$ electron
in the $x'y'z'$ coordinate.

\subsection{$E$1 transition at the $L_{2,3}$ edge}

The Hamiltonian of photon may be written as
\begin{equation}
 H_{\textrm{ph}} = \sum_{{\bf q}, \mu}\omega_{\bf q}
              c_{{\bf q}\mu}^{\dagger}c_{{\bf q}\mu},
\end{equation}
where $c_{\textbf{q}\mu}$ ($c_{\textbf{q} \mu}^{\dagger}$)
stands for the annihilation (creation) 
operator of the photon with momentum $\textbf{q}$,
energy $\omega_{\textbf{q}}$, and polarization direction
$\mu$ ($=x,y$, and $z$).
In the electric dipole ($E$1) 
transition, a $2p$-core electron is excited to 
the $3d$ states at the transition-metal $L_{2,3}$ edge.
The $2p$ states are characterized by the 
total angular momentum $j=3/2$ and $1/2$
due to the strong spin-orbit interaction. 
The eigenstates with $j=3/2$ may be expressed as
$|\phi_{1}\uparrow\rangle$, 
$\sqrt{1/3}|\phi_{1}\downarrow\rangle+\sqrt{2/3}|\phi_{0}\uparrow\rangle$, 
$\sqrt{2/3}|\phi_{0}\downarrow\rangle+\sqrt{1/3}|\phi_{-1}\uparrow\rangle$,
$|\phi_{-1}\downarrow\rangle$, for $m=3/2$, $1/2$, $-1/2$, $-3/2$, 
respectively, and those with $j=1/2$ may be expressed as
$-\sqrt{2/3}|\phi_{1}\downarrow\rangle+\sqrt{1/3}|\phi_{0}\uparrow\rangle$, 
$-\sqrt{1/3}|\phi_{0}\downarrow\rangle+\sqrt{2/3}|\phi_{-1}\uparrow\rangle$, 
for $m=1/2$ and $-1/2$, respectively,
where $m$ represents the magnetic quantum number. 
The orbitals $\phi_{1}$, $\phi_{0}$, and $\phi_{-1}$ 
have angular dependence $Y_{11}$, $Y_{10}$ and $Y_{1-1}$, respectively. 
In the above expression, the coordinate frame for spin 
is defined the same as the orbitals, 
which is different from that for the spins of the $3d$ states, 
that is, $\uparrow$ and $\downarrow$ are 
associated with the direction of the crystal $c$ axis.
Therefore, the interaction between photon and electron at site $i$ 
may be described as
\begin{equation}
H_{\rm int}=w\sum_{\textbf{q}, \mu}
\frac{1}{\sqrt{2\omega_{\textbf{q}}}}
\sum_{i, m, \sigma}D^{\mu}(jm,\sigma)
h_{jm}^{\dagger}c_{\textbf{q}\mu}d_{i\sigma}
{\rm e}^{i\textbf{q}\cdot\textbf{r}_{i}} +{\rm H.c.},
\label{eq.tran1}
\end{equation}
where $h_{jm}^{\dagger}$ stands for the creation operator of the $2p$ hole with
the angular momentum $jm$. 
The $w$ is a constant proportional to 
$\int_0^{\infty}r^3R_{3d}(r)R_{2p}(r){\rm d}r$,
with $R_{3d}(r)$ and $R_{2p}(r)$ being the radial wave-functions for 
the $3d$ and $2p$ states of Cu atom. The $D^{\mu}(jm,\sigma)$ describes
the dependence on the core-hole state and $3d$ spin, which is calculated
by taking care of the difference in the definition of the spin axes
between the $3d$ and $2p$ states. 
Table \ref{table.1} shows the calculated values of
$D^{\mu}(jm,\sigma)$.
Note that no $E1$ transition takes place for the polarization parallel
to the $z$ axis. 

\begin{table}
\caption{\label{table.1}
Numerical values of $D^{\mu}(jm,\sigma)$. Here, 
$u={\rm e}^{i(\alpha+\gamma)/2}\cos\frac{\beta}{2}$ and 
$v={\rm e}^{i(\gamma-\alpha)/2}\sin\frac{\beta}{2}$,
where $\alpha$, $\beta$, and $\gamma$ are the
Euler angles transforming the 
crystal-fixed $a$, $b$, and $c$ axes to the spin axes $x'$, $y'$, 
and $z'$ axes.}
\begin{ruledtabular}
\begin{tabular}{rrrrrr}
$\mu$ & $J$   & $m$   & $\sigma=\uparrow$ & $\sigma=\downarrow$ \\
\hline
$x$ & $3/2$ & $3/2$  & $-\frac{1}{\sqrt{10}}u^{*}$ & $\frac{1}{\sqrt{10}}v$ \\ 
    &       & $1/2$  & $-\frac{1}{\sqrt{30}}v^{*}$
                     & $-\frac{1}{\sqrt{30}}u$ \\
    &       & $-1/2$ & $\frac{1}{\sqrt{30}}u^{*}$
                     & $-\frac{1}{\sqrt{30}}v$               \\
    &       & $-3/2$ & $\frac{1}{\sqrt{10}}v^{*}$ 
                     & $\frac{1}{\sqrt{10}}u$ \\
    & $1/2$ & $1/2$  & $\frac{1}{\sqrt{15}}v^{*}$
                     & $\frac{1}{\sqrt{15}}u$ \\
    &       & $-1/2$ & $\frac{1}{\sqrt{15}}u^{*}$
                     & $-\frac{1}{\sqrt{15}}v$ \\
\hline
$y$ & $3/2$ & $3/2$  & $\frac{i}{\sqrt{10}}u^{*}$ & $-\frac{i}{\sqrt{10}}v$ \\ 
    &       & $1/2$  & $\frac{i}{\sqrt{30}}v^{*}$
                     & $\frac{i}{\sqrt{30}}u$ \\
    &       & $-1/2$ & $\frac{i}{\sqrt{30}}u^{*}$
                     & $-\frac{i}{\sqrt{30}}v$                \\
    &       & $-3/2$ & $\frac{i}{\sqrt{10}}v^{*}$ 
                     & $\frac{i}{\sqrt{10}}u$ \\
    & $1/2$ & $1/2$  & $-\frac{i}{\sqrt{15}}v^{*}$
                     & $-\frac{i}{\sqrt{15}}u$ \\
    &       & $-1/2$ & $\frac{i}{\sqrt{15}}u^{*}$
                     & $-\frac{i}{\sqrt{15}}v$ \\

\end{tabular}
\end{ruledtabular}
\end{table}

\subsection{Formulation of RIXS spectra}

Following Nozi\'{e}res and Abrahams,\cite{Nozieres74} 
we use the Keldysh-Schwinger formalism\cite{Keldysh65}
to investigate the RIXS spectra. First we prepare the initial state
\begin{equation}
 |\Phi_i\rangle = c_{{\bf q}_i\alpha_i}^{\dagger}|g\rangle,
\end{equation}
where $|g\rangle$ represent the ground state of the matter with energy 
$E_g$.
The incident photon has momentum $\textbf{q}_i$,
energy $\omega_i$, and polarization direction $\alpha_i$.
Then we calculate the probability of finding a photon with
momentum $\textbf{q}_f$, energy $\omega_f$, and polarization direction
$\alpha_f$ at time $t_0$ using the following formula,
\begin{equation}
 P_{{\bf q}_f\alpha_f;{\bf q}_i\alpha_i}(t_0)
  =\langle \Phi_i| U(-\infty,t_0)c_{{\bf q}_f\alpha_f}^\dagger
              c_{{\bf q}_f\alpha_f}U(t_0,-\infty) |\Phi_i\rangle .
\label{eq.prob}
\end{equation}
Here $U(t,t')$ is the $S$ matrix defined by
\begin{equation}
 U(t,-\infty) = T\exp\left\{-i\int_{-\infty}^t H_{\rm int}(t'){\rm d}t'\right\},
\end{equation}
where $H_{\rm int}(t)=\exp[i(H+H_{\rm ph})t] H_{\rm int}
\exp[-i(H+H_{\rm ph})t]$ with $H$ and $H_{\rm ph}$ being the Hamiltonian of 
matter and photon, and $T$ represents the time-ordering operator.

The S-matrix is expanded up to the second order with $H_{\rm int}$:
\begin{eqnarray}
 U(t,-\infty) &=& 1 + (-i)\int_{-\infty}^t H_{\rm int}(t'){\rm d}t' \nonumber \\
              &+& (-i)^2\int_{-\infty}^t {\rm d}t'\int_{-\infty}^{t'}{\rm d}t''
	        H_{\rm int}(t')H_{\rm int}(t'') .
\nonumber \\
\end{eqnarray}
By substituting this into Eq.~(\ref{eq.prob}),
we obtain the transition probability per unit time with $t_{0}\to\infty$,
$W(q_f\alpha_f;q_i\alpha_i)$ in the following form:
\begin{eqnarray}
 & & W(q_f\alpha_f;q_i\alpha_i) \nonumber \\
 &=& \int_{-\infty}^{0}{\rm d}t
 \int_{-\infty}^{\infty}{\rm d}u'\int_{-\infty}^{u'}{\rm d}t'\sum_{f'}
 \sum_{n, n'}  \langle\Phi_i|H_{\rm int}(t')|n'\rangle \nonumber\\
 && \times
\langle n'|H_{\rm int}(u')|\Phi_{f'}
 \rangle \langle \Phi_{f'}|H_{\rm int}(u)|n\rangle\langle n|H_{\rm int}(t)|
 \Phi_i\rangle 
     \nonumber\\
 &=& 2\pi\sum_{f'}\left|\sum_{n}\langle f|H_{\rm int}|n\rangle
  \frac{1}{E_g+\omega_i-E_n} \langle n|H_{\rm int}
 |\Phi_i\rangle\right|^2 \nonumber\\
 &\times&\delta(E_g+\omega_i-E_{f'}-\omega_f),
\label{eq.optical}
\end{eqnarray} 
with $q_i\equiv({\bf q}_i,\omega_i)$, $q_f\equiv({\bf q}_f,\omega_f)$, and
$|\Phi_{f'}\rangle=c_{q_f \alpha_f}|f'\rangle$.
This is nothing but conventional expression of the second-order dipole allowed 
process, on which our following analysis is based.
The $|f'\rangle$ represents the eigenstate of the Hamiltonian of the matter 
with the eigenvalue $E_{f'}$, while $|n\rangle$ represents the eigenstate
with eigenvalue $E_n$ in the presence of core hole.

In our previous paper having discussed RIXS at the Cu $K$ 
edge,\cite{Nagao07} 
the $E1$ transition could not directly change the $3d$ states
but simply create the core-hole potential which attracts electrons in the
$3d$ states. In such a situation, by 
treating the core-hole potential as a
perturbation, we could conveniently introduce the Green 
functions and diagrams on the basis of the Keldysh-Schwinger formalism
to calculate $W(q_f\alpha_f;q_i\alpha_i)$.
To the present case, however, such diagrammatic procedure is difficult to apply,
since the $E1$ transition modifies the $3d$ states from the 
$3d^9$-configuration to the $3d^{10}$-configuration, which change is hard
to express by the diagrams for magnon excitations.

\section{\label{sect.3}Magnon excitations around the core-hole site}

\subsection{General consideration}
\begin{figure}
\includegraphics[width=8.0cm]{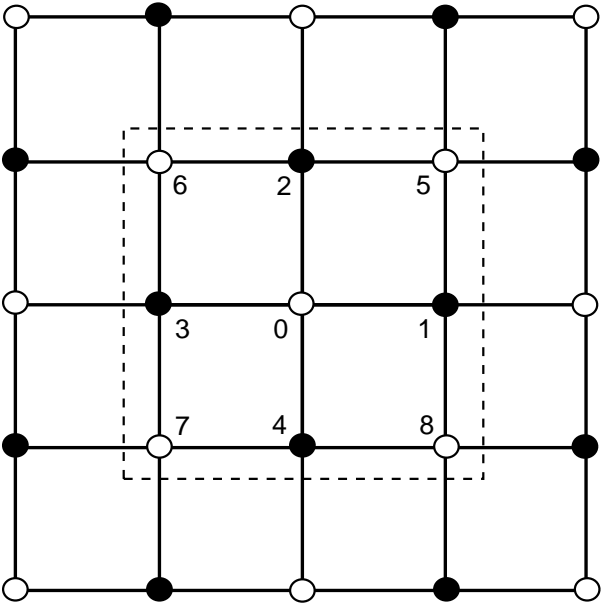}%
\caption{\label{fig.cluster}
Spins near the core-hole. 
Each site is labeled by number and the core-hole site is
assigned number 0.
}
\end{figure}

Assuming that the core hole is created at the origin in the intermediate
state, we analyze the spin system around the origin, as shown 
in Fig.~\ref{fig.cluster}.
We write the ground state $|g\rangle$ of $H_{\rm mag}$ as
\begin{equation}
 |g\rangle = |\uparrow\rangle|\psi_0^{\uparrow}\rangle
           + |\downarrow\rangle|\psi_0^{\downarrow}\rangle,
\end{equation}
where $|\uparrow\rangle$ and $|\downarrow\rangle$ represent the spin
states at the origin, and $|\psi_0^{\uparrow}\rangle$ and 
$|\psi_0^{\downarrow}\rangle$ are constructed 
by the bases of the rest of 
spins. The expectation value of $S^{z}$ at the origin
may be expressed as 
\begin{equation}
 \langle S_{0}^{z'}\rangle = 
 \frac{1}{2}\left\{ \langle\psi_0^{\uparrow}|\psi_0^{\uparrow}\rangle
                   -\langle\psi_0^{\downarrow}|\psi_0^{\downarrow}\rangle
            \right\}.
\label{eq.magnetization}
\end{equation}

Since the spin degree of freedom is lost at the origin just after the $E1$ 
transition takes place, the wave function becomes
\begin{equation}
 H_{\rm int}|g\rangle \propto\sum_{m}\left[ 
\sum_{\sigma=\uparrow,\downarrow}
    D^{\alpha_i}(jm,\sigma)|\psi_0^{\sigma}\rangle
  \right] |jm\rangle,
\end{equation}
where $|jm\rangle$ represents the core hole state.
Let $H'$ be the Hamiltonian in the intermediate state.
It consists of the exchange interaction terms excluding those 
with the core-hole site. The Hilbert space representing $H'$ is different from
the initial state. This situation is the same as the problem of 
the non-magnetic impurity.\cite{Tonegawa68,Wan93} 
Note that the states $|\psi_0^{\uparrow}\rangle$ and 
$|\psi_0^{\downarrow}\rangle$ are not 
the eigenstates of $H'$. They are expanded in terms of 
normalized eigenstate $|\phi_{\eta}\rangle$'s of $H'$ with eigenvalue $\epsilon'_{\eta}$. 
We therefore have
\begin{eqnarray}
 &&\sum_n H_{\rm int}|n\rangle\frac{1}{\omega_i+E_g-E_n}
   \langle n|H_{\rm int}|\Phi_i\rangle \nonumber \\
   &\propto& \sum_{m, \sigma, \sigma'} D^{\alpha_f}(jm,\sigma)^{*}
                             D^{\alpha_i}(jm,\sigma') \nonumber\\
&\times&
   \sum_{\eta}|\sigma\rangle |\phi_{\eta}\rangle R(\epsilon'_{\eta})
    \langle\phi_{\eta}|\psi_{0}^{\sigma'}\rangle,
\label{eq.process1}
\end{eqnarray}
with
\begin{equation}
 R(\epsilon'_{\eta}) =
 \frac{1}{\omega_i+\epsilon_g -\epsilon_{\rm core}+i\Gamma - \epsilon'_{\eta}},
\end{equation}
where $\epsilon_g$ represents the ground state energy of $H_{\rm mag}$.
The $\epsilon_{\rm core}$ denotes 
the energy required to create a core hole in the state 
$|jm\rangle$ and the $3d^{10}$-configuration.
The $\Gamma$ stands for the life-time broadening width of the core hole;
$\Gamma\sim 0.3$ eV at the $L_3$ edge.
Since $H'$ is different from $H_{\rm mag}$ only around the core-hole site, 
$|\phi_{\eta}\rangle$ could have sufficient overlap
with $|\psi_0^{\sigma'}\rangle$ only when $\epsilon'_{\eta}-\epsilon_g$ varies 
in the range of several $J$'s ($J\sim 0.1$ eV).
Note that the spin degree of freedom at the origin is recovered in the final
state and the wave function is given by $|\sigma\rangle|\phi_{\eta}\rangle$. 
The first factor in Eq.~(\ref{eq.process1}) has relations,
\begin{eqnarray}
\sum_{m}D^{\alpha_f}(jm,\sigma)^{*}D^{\alpha_i}(jm,\sigma)
&\equiv& P_{\sigma}^{(0)}(j;\alpha_f,\alpha_i), \\
\sum_{m}D^{\alpha_f}(jm,\sigma)^{*}D^{\alpha_i}(jm,-\sigma) 
&\equiv& P_{\sigma}^{(1)}(j;\alpha_f,\alpha_i).
\end{eqnarray}
Here, for $\sigma=\uparrow$, $-\sigma$ denotes $\downarrow$, and vice versa. 
Table \ref{table.2} shows $P_{\sigma}^{(0)}$ and $P_{\sigma}^{(1)}$ 
for $\alpha_i$ and $\alpha_f$ along the $x$, $y$, and $z$ axes.
The extension to the cases of general $\alpha_i$ and $\alpha_f$ directions
is obvious. The $P_{\sigma}^{(0)}$ and $P_{\sigma}^{(1)}$ correspond to 
the spin-conserving and the spin-flip processes, respectively.
Note that 
$P_{\sigma}^{(1)}(1/2;\alpha_f,\alpha_i)
+P_{\sigma}^{(1)}(3/2;\alpha_f,\alpha_i)=0$.
As will become clear later, spin-flip excitation spectra are proportional to
$|P^{(1)}|^2$. Therefore, if the processes for $j=3/2$ and $j=1/2$ are not 
separated, no spin-flip excitation comes out. 

\begin{table*}
\caption{\label{table.2}
$P_{\sigma}^{(0)}(j;\alpha_f,\alpha_i)$ and 
$P_{\sigma}^{(1)}(j;\alpha_f,\alpha_i)$ 
where upper and lower signs correspond to $\sigma=\uparrow$ and $\downarrow$, 
respectively.
}
\begin{ruledtabular}
\begin{tabular}{rcrrrrrrr}
      & & $P_{\sigma}^{(0)}$ & & & & $P_{\sigma}^{(1)}$ & & \\
\hline
$j$   & $\alpha_f \setminus \alpha_i$ & $x$  & $y$ & $z$ &  & $x$ & $y$ & $z$ \\ 
\hline
$\frac{3}{2}$ & $x$ & $\frac{2}{15}$ & $\mp \frac{i}{15}\cos\beta$ & $0$ & 
& $0$ & $\frac{i}{15} \textrm{e}^{\pm i \gamma} \sin\beta$ & $0$ \\
      & $y$ & $\pm \frac{i}{15}\cos\beta$ & $\frac{2}{15}$ & $0$ & & $-\frac{i}{15} \textrm{e}^{\pm i \gamma} \sin\beta$ & $0$ & $0$ \\
      & $z$ & $0$ & $0$ & $0$ & & $0$ & $0$ & $0$ \\
\hline
$\frac{1}{2}$ & $x$ & $\frac{1}{15}$ & $\pm \frac{i}{15}\cos\beta$ 
& $0$ & & $0$ & 
$-\frac{i}{15} \textrm{e}^{\pm i \gamma} \sin\beta$ & $0$ \\ 
      & $y$ & $\mp \frac{i}{15}\cos\beta$
 & $\frac{1}{15}$ & $0$ & & 
$\frac{i}{15} \textrm{e}^{\pm i \gamma} \sin\beta$ & $0$ & $0$ \\
      & $z$ & $0$ & $0$ & $0$ & & $0$ & $0$ & $0$ \\
\end{tabular}
\end{ruledtabular}
\end{table*}

\subsection{Scattering channel with changing the polarization}

We analyze the scattering channel that the polarization changes;
let $\alpha_i$ and $\alpha_f$ be along $y$ and $x$ axes, respectively.  
We assume that the staggered magnetic moment is parallel to the z'-axis, where 
the coordinate frame of the $x',y',z'$ axes
is specified by the Euler angles $\alpha$, 
$\beta$, and $\gamma$ from the coordinate frame of the $x$, $y$, $z$ axes.
We have the spin-conserving term coming from $P_{\sigma}^{(0)}$ and
the spin-flipping term coming from $P_{\sigma}^{(1)}$.
The spin-conserving term is given by
\begin{eqnarray}
 &&\sum_n H_{\rm int}|n\rangle\frac{1}{\omega_i+E_g-E_n}
   \langle n|H_{\rm int}|g\rangle \nonumber \\
 &\propto& P_{\uparrow}^{(0)}
  |\uparrow\rangle\sum_{\eta}|\phi_{\eta}\rangle R(\epsilon'_{\eta})
  \langle\phi_{\eta}|\psi_{0}^{\uparrow}\rangle  \nonumber \\
& +& P_{\downarrow}^{(0)}
  |\downarrow\rangle\sum_{\eta}|\phi_{\eta}\rangle R(\epsilon'_{\eta})
  \langle\phi_{\eta}|\psi_{0}^{\downarrow}\rangle, \nonumber\\
 &\propto& \left(-\frac{i}{15}\right)\cos\beta \left\{
  |\uparrow\rangle\sum_{\eta}|\phi_{\eta}\rangle R(\epsilon'_{\eta})
  \langle\phi_{\eta}|\psi_{0}^{\uparrow}\rangle \right. \nonumber \\
  & & - \left. 
  |\downarrow\rangle\sum_{\eta}|\phi_{\eta}\rangle R(\epsilon'_{\eta})
  \langle\phi_{\eta}|\psi_{0}^{\downarrow}\rangle \right\}. 
\label{eq.non-flip}
\end{eqnarray}
Since this state has an overlap to the ground state, we have
the elastic amplitude $A_{\rm elas}$, 
\begin{eqnarray}
 A_{\rm elas} &\propto& \left(-\frac{i}{15}\right)\cos\beta \left\{
  \langle\psi_0^{\uparrow}|\sum_{\eta}|\phi_{\eta}\rangle R(\epsilon'_{\eta})
   \langle\phi_{\eta}|\psi_0^{\uparrow}\rangle \right. \nonumber \\
  & & - \left. 
\langle\psi_0^{\downarrow}|\sum_{\eta}|\phi_{\eta}\rangle R(\epsilon'_{\eta})
   \langle\phi_{\eta}|\psi_0^{\downarrow}\rangle
      \right\}.
\label{eq.elastic}
\end{eqnarray}

Introducing the quantity 
\begin{equation}
   f_{\sigma}^{(1)}(\omega_i) =
    \frac{1}{\langle\psi_{0}^{\sigma}|\psi_{0}^{\sigma}\rangle}
     \langle\psi_0^{\sigma}| \sum_{\eta}
     |\phi_{\eta}\rangle R(\epsilon_{\eta}')
     \langle\phi_{\eta}|\psi_0^{\sigma}\rangle, 
\label{eq.B}
\end{equation}
with $\sigma=\uparrow$ and $\downarrow$, we define $f_{0}^{(1)}(\omega_i)$
and $\Delta(\omega_i)$ by
\begin{equation}
   f_{\sigma}^{(1)}(\omega_i) = f_{0}^{(1)}(\omega_i)
   \pm \frac{1}{2}\Delta(\omega_i),
\label{eq.BD}
\end{equation}
where plus and minus signs in the second term correspond to
$\sigma=\uparrow$ and $\downarrow$, respectively.
In the far-off-resonance condition that $\omega_i\ll\epsilon_{\rm core}$
or $\omega_i\gg\epsilon_{\rm core}$, and in the large limit of $\Gamma$, 
which is called as the UCL condition, that  
$\Gamma\gg |\omega_i-\epsilon_{\rm core}|, |\epsilon'_{\eta}-\epsilon_g|$,
we could factor out $R(\epsilon'_{\eta})$ 
from the summation over $\eta$
in Eq.~(\ref{eq.B}). 
Thereby, using $\sum_{\eta}|\phi_{\eta}\rangle\langle\phi_{\eta}|=1$,
we immediately obtain $\Delta(\omega_i)=0$.

By inserting Eq.~(\ref{eq.BD}) in Eq.~(\ref{eq.elastic}), 
and by using Eq. (\ref{eq.magnetization}), we have
\begin{equation}
 A_{\rm elas} \propto \left(-\frac{i}{15}\right)
 \cos\beta \left\{2f_{0}^{(1)}(\omega_i)
 \langle S_0^{z'}\rangle
              +\frac{1}{2}\Delta(\omega_i)\right\}.
\label{eq.elasDel}
\end{equation}
This result may be compared with the conventional expression of 
the elastic magnetic scattering amplitude by Hannon {\it et al.},\cite{Hannon1988}
\begin{equation}
\sigma^{(1)} \mbox{\boldmath{$\alpha$}}_f\times
\mbox{\boldmath{$\alpha$}}_i\cdot{\bf m},
\label{eq.Hannon}
\end{equation}
where ${\bf m}$ is the staggered magnetic moment vector, and
$\sigma^{(1)}$ is a certain numerical constant.
In order that Eq.~(\ref{eq.elasDel}) is consistent with Eq.~(\ref{eq.Hannon}),
$\Delta(\omega_i)$ could be, if it existed, expanded as 
$a\sum_{\delta}\langle S_{\delta}^{z'}\rangle+b\sum_{\delta'}\langle 
S_{\delta'}^{z'}\rangle+\cdots$ with $\delta$ and $\delta'$ denoting the 
nearest and next nearest neighbor sites to the core-hole site $0$ 
($a$ and $b$ have to go to zero
in the far-off-resonance condition as well as in the UCL condition),
since $\Delta(\omega_i)$ should be incorporated into the renormalization
of $\sigma^{(1)}$ in Eq.~(\ref{eq.Hannon}).
The presence of $\Delta(\omega_i)$ may suggest that the disturbance through 
the intermediate state reaches to the neighboring lattice sites.
Actually, Eq. (\ref{eq.non-flip}) has finite overlaps to 
$S_{\delta}^{z'} |g\rangle$ and $S_{\delta'}^{z'} |g\rangle$, which
are not orthogonal to both 
$S_{0}^{z'} |g\rangle$ and $S_{0}^{z'} |g\rangle$.
By neglecting such possibility, 
we project Eq.~(\ref{eq.non-flip}) onto $|g\rangle$ and 
$S_{0}^{z'}|g\rangle$. Since $S_{0}^{z'}|g\rangle$ is not orthogonal to 
$|g\rangle$ and not normalized, we need to introduce overlap matrix 
$\hat{\rho}$ between $|g\rangle$ and $S_{0}^{z'}|g\rangle$. 
Using the inverse of $\hat{\rho}$,  we have
\begin{eqnarray}
 &&\sum_n H_{\rm int}|n\rangle\frac{1}{\omega_i+E_g-E_n}
   \langle n|H_{\rm int}|g\rangle \nonumber\\
 &=& \left(-\frac{i}{15}\right)
 \cos\beta \left\{
 \frac{1}{2}\Delta(\omega_i)|g\rangle
 +2f_{0}^{(1)}(\omega_i) S_0^{z'}|g\rangle
	      \right\}.
\end{eqnarray}
By neglecting $\Delta(\omega_i)$, the final state
in this channel is approximated as 
\begin{eqnarray}
& & \left(-\frac{i}{15}\right)\cos\beta 2f_{0}^{(1)}(\omega_i) S_0^{z'}|g\rangle
\nonumber \\
&=& \left(-\frac{i}{15}\right)2f_{0}^{(1)}(\omega_i)
\mbox{\boldmath{$\alpha$}}_{f\perp}\times
 \mbox{\boldmath{$\alpha$}}_{i\perp}\cdot {\bf S}_{0\parallel}|g\rangle,
\label{eq.sparallel}
\end{eqnarray}
where ${\bf S}_{0\parallel}$ stands for the component of ${\bf S}_0$ 
parallel to the direction of the staggered magnetic moment.
The $\mbox{\boldmath{$\alpha$}}_{i\perp}$ and 
$\mbox{\boldmath{$\alpha$}}_{f\perp}$ represent the polarization vectors
projected onto the $x$-$y$ plane. Therefore 
$\mbox{\boldmath{$\alpha$}}_{f\perp}\times
\mbox{\boldmath{$\alpha$}}_{i\perp}$ is always parallel to the $z$ axis.
Note that the inelastic terms are sometimes inferred from 
Eq.~(\ref{eq.Hannon}) with simply replacing ${\bf m}$ by the spin operator 
${\bf S}_0$ at site 0.\cite{Haverkort2010} 
Equation (\ref{eq.sparallel}) is, however, different from such a term, 
since Eq.~(\ref{eq.sparallel}) is restricted within the spin-conserving process,
and disappears for $\beta=\pi/2$.

The final states responsible to inelastic scattering mainly come 
from the spin-flip terms, 
\begin{eqnarray}
  &&\sum_n H_{\rm int}|n\rangle\frac{1}{\omega_i+E_g-E_n}
   \langle n|H_{\rm int}|\Phi_i\rangle \nonumber \\
  &\propto& P_{\downarrow}^{(1)}
  |\downarrow\rangle\sum_{\eta}|\phi_{\eta}\rangle R(\epsilon'_{\eta})
  \langle\phi_{\eta}|\psi_{0}^{\downarrow}\rangle \nonumber \\
  &+& P_{\uparrow}^{(1)}
  |\uparrow\rangle\sum_{\eta}|\phi_{\eta}\rangle R(\epsilon'_{\eta})
  \langle\phi_{\eta}|\psi_{0}^{\uparrow}\rangle, \nonumber\\
  &\propto& \left(\frac{i}{15}\right)
  \sin\beta \left\{{\rm e}^{-i\gamma}
   |\downarrow\rangle\sum_{\eta}|\phi_{\eta}\rangle R(\epsilon'_{\eta})
   \langle\phi_{\eta}|\psi_{0}^{\uparrow}\rangle \right. \nonumber \\
   & & + \left.
   {\rm e}^{i\gamma}|\uparrow\rangle\sum_{\eta}|\phi_{\eta}\rangle
  R(\epsilon'_{\eta})\langle\phi_{\eta}|\psi_{0}^{\downarrow}\rangle \right\}.
\label{eq.spin-flip}
\end{eqnarray}
We project this state onto $S_{0}^{\pm}|g\rangle$ with
$S_0^{\pm}=S_0^{x'}\pm i S_0^{y'}$. These states are orthogonal to each other 
and to $|g\rangle$ but not normalized, that is,
$\langle g|S_0^{-}S_0^{+}|g\rangle=
\langle\psi_0^{\downarrow}|\psi_0^{\downarrow}\rangle$ and
$\langle g|S_0^{+}S_0^{-}|g\rangle=
\langle\psi_0^{\uparrow}|\psi_0^{\uparrow}\rangle$.
Thereby we could express Eq.~(\ref{eq.spin-flip}) as
\begin{eqnarray}
 &\propto& \left(\frac{i}{15}\right)
 \sin\beta \Bigl\{
 f_{0}^{(1)}(\omega_i)
 ({\rm e}^{-i\gamma}S_0^{-}+{\rm e}^{i\gamma}S_0^{+})|g\rangle
 \nonumber\\
 & & + \frac{1}{2}\Delta(\omega_i)
 ({\rm e}^{-i\gamma}S_0^{-}-{\rm e}^{i\gamma}S_0^{+})|g\rangle
 \Bigr\} .
\label{eq.sperp}
\end{eqnarray}
The first term may be rewritten as
\begin{eqnarray}
 & & \left(\frac{i}{15}\right) \sin\beta 2f_{0}^{(1)}(\omega_i)
  (\cos\gamma S_{0}^{x'}-\sin\gamma S_{0}^{y'})|g\rangle \nonumber\\
  &=& \left(-\frac{i}{15}\right) 2f_{0}^{(1)}(\omega_i)
  \mbox{\boldmath{$\alpha$}}_{f\perp}\times 
  \mbox{\boldmath{$\alpha$}}_{i\perp} \cdot {\bf S}_{0\perp}|g\rangle,
\end{eqnarray}
where ${\bf S}_{0\perp}$ represents the component perpendicular to
the direction of the staggered magnetic moment.
As regards the second term of Eq.~(\ref{eq.sperp}),
the inclusion of $\Delta(\omega_i)$ would require 
adding the states $S_{\delta}^{\pm}|g\rangle$ and $S_{\delta'}^{\pm}|g\rangle$
as the states to be projected.
Such an analysis would be rather complicated and will not be attempted
in this paper.
With disregarding the term of $\Delta(\omega_i)$, we finally have 
the expression of the final state by combining Eq.~(\ref{eq.sparallel}) 
with Eq.~(\ref{eq.sperp});
\begin{equation}
  \sum_n H_{\rm int}|n\rangle\frac{1}{\omega_i+E_g-E_n}
   \langle n|H_{\rm int}|g\rangle 
   =2f_{0}^{(1)}(\omega_i)
    \mbox{\boldmath{$\alpha$}}_f\times \mbox{\boldmath{$\alpha$}}_i
    \cdot {\bf S}_{0}|g\rangle.
\end{equation}

\subsection{Scattering channel without changing the polarization}

In this scattering channel, only the spin-conserving excitations are brought 
about through the diagonal components of $P_{\sigma}^{(0)}$. 
For $\mbox{\boldmath{$\alpha$}}_i=\mbox{\boldmath{$\alpha$}}_f=(1,0,0)$
and for $\mbox{\boldmath{$\alpha$}}_i=\mbox{\boldmath{$\alpha$}}_f=(0,1,0)$,
we have
\begin{eqnarray}
 &&\sum_n H_{\rm int}|n\rangle\frac{1}{\omega_i+E_g-E_n}
   \langle n|H_{\rm int}|g\rangle \nonumber \\
 &\propto& \left(\frac{2}{15}\right) \left\{
  |\uparrow\rangle\sum_{\eta}|\phi_{\eta}\rangle R(\epsilon'_{\eta})
  \langle\phi_{\eta}|\psi_{0}^{\uparrow}\rangle \right. \nonumber \\
  & & \left.
  + |\downarrow\rangle\sum_{\eta}|\phi_{\eta}\rangle R(\epsilon'_{\eta})
  \langle\phi_{\eta}|\psi_{0}^{\downarrow}\rangle \right\}. 
\label{eq.diago}
\end{eqnarray}

We consider the spherical form of the final state,
${\bf X}\cdot {\bf S}_0|g\rangle$ with 
${\bf X}\equiv \sum_{\delta}{\bf S}_{\delta}$.
The deviation from the spherical form would be considered
as a next-step approximation, since we are neglecting $\Delta(\omega_i)$
which suggests that ${\bf S}_{\delta}|g\rangle$ are to be included
as possible excited states.
Since ${\bf X}\cdot {\bf S}_0|g\rangle$ is not orthogonal to $|g\rangle$,
we introduce the overlap matrix $\hat{\rho}$
defined by $(\hat{\rho})_{i,j}\equiv \langle\psi_i|\psi_j\rangle$
with $|\psi_1\rangle\equiv |g\rangle$ and 
$|\psi_2\rangle\equiv {\bf X}\cdot{\bf S}_0|g\rangle$.
Then, using the inverse of $\hat{\rho}$,
we project the final state onto these states, resulting that
\begin{eqnarray}
 &&\sum_n H_{\rm int}|n\rangle\frac{1}{\omega_i+E_g-E_n}
   \langle n|H_{\rm int}|g\rangle \nonumber \\
&\propto& 
    f_{1}^{(2)}(\omega_i)|g\rangle 
  + f_{2}^{(2)}(\omega_i){\bf X}\cdot{\bf S}_0|g\rangle
\label{eq.spin-conserve}
\end{eqnarray}
where $f_{m}^{(2)}(\omega_i)$'s are given by
\begin{equation}
 f_{m}^{(2)}(\omega_i) = \sum_{m'}
     (\hat{\rho}^{-1})_{m,m'}Q_{m'}^{(2)}(\omega_i),
\end{equation}
with
\begin{eqnarray}
 Q_{1}^{(2)}(\omega_i) &=&
   \sum_{\sigma} \langle\psi_0^{\sigma}| \sum_{\eta}
   |\phi_{\eta}\rangle R(\epsilon'_{\eta})
   \langle\phi_{\eta}|\psi_0^{\sigma}\rangle, \label{eq.f21}\\
 Q_{2}^{(2)}(\omega_i) &=&
   \frac{1}{2}\Bigl\{\sum_{\sigma} \textrm{sgn}(\sigma)
  \langle\psi_0^{\sigma}|X^{z'} \sum_{\eta}
   |\phi_{\eta}\rangle R(\epsilon'_{\eta})
   \langle\phi_{\eta}|\psi_0^{\sigma}\rangle \nonumber \\
          & & + \langle\psi_0^{\uparrow}|X^{-} \sum_{\eta}
    |\phi_{\eta}\rangle R(\epsilon'_{\eta})
    \langle\phi_{\eta}|\psi_0^{\downarrow}\rangle \nonumber \\
  & & + \langle\psi_0^{\downarrow}|X^{+} \sum_{\eta}
    |\phi_{\eta}\rangle R(\epsilon'_{\eta})
    \langle\phi_{\eta}|\psi_0^{\uparrow}\rangle \Bigr\},
\label{eq.f22}
\end{eqnarray}
where $\textrm{sgn}(\sigma)=1$ for $\sigma=\uparrow$ and $-1$ for 
$\sigma=\downarrow$.

To understand the possible spin-flip at the origin in 
Eq.~(\ref{eq.spin-conserve}), we take up the second term
of Eq.~(\ref{eq.f22}), and examine the associated generating process.
Consider that the down spin at the core-hole site is annihilated by
absorbing photon. The wave function for surrounding spins is expressed by 
$|\psi_{0}^{\downarrow}\rangle$, which satisfies 
$\langle\psi_{0}^{\downarrow}|(\sum_{i}S_i^{z})|\psi_{0}^{\downarrow}
\rangle=1/2$ with $i$ running all the lattice sites except the origin.
At the end of time evolution in the intermediate state, 
this wave function is modified, and could have a finite overlap
with the state $X^{+}|\psi_{0}^{\uparrow}\rangle$, since
$\langle\psi_{0}^{\uparrow}|X^{-}(\sum_{i}S_i^{z'})
X^{+}|\psi_{0}^{\uparrow}\rangle=1/2$.
Then, the $3d$ hole
with down spin is created at the core-hole site by annihilating the core hole
with emitting photon, leading to 
$|\downarrow\rangle X^{+}|\psi_{0}^{\uparrow}\rangle$. Since
$S_{0}^{-}|g\rangle=|\downarrow\rangle|\psi_{0}^{\uparrow}\rangle$, 
this state is nothing but $X^{+}S_{0}^{-}|g\rangle$.

In the far-off-resonance condition and in the UCL approximation, 
$R(\epsilon'_{\eta})$ could be factored out in Eqs.~(\ref{eq.f21}) and
(\ref{eq.f22}). It leads to
$Q_{m}(\omega_i)= R(\epsilon'_{0})(\hat{\rho})_{m,1}$ and as a consequence, 
\begin{equation}
 f_{m}^{(2)}(\omega_i)= R(\epsilon'_{0})
 \sum_{m'}(\hat{\rho}^{-1})_{m,m'}(\hat{\rho})_{m',1}
                      =R(\epsilon'_{0})\delta_{m,1}.
\end{equation}
Therefore, in both cases, the spin-conserving excitations could not be 
generated.

One may formally write the intermediate state 
Hamiltonian $H'$ by eliminating 
the bonds which connect the spin at the origin 
to spins at neighboring sites from the initial state Hamiltonian
$H_{\rm mag}$:
\begin{equation}
 H' = H_{\rm mag} + V , \quad 
 V = -J\sum_{\delta}{\bf S}_{0}\cdot{\bf S}_{\delta}.
\label{eq.V}
\end{equation}
The Hilbert space representing Eq.~(\ref{eq.V}) formally contains 
the spin-degrees of freedom at the core-hole site, which should be
completely decoupled from the outer spins in the final solution,
as known from the non-magnetic impurity problem.\cite{Tonegawa68,Wan93}
Finite-order perturbation with $V$ could not satisfy this criteria.
In this context, although a first-order perturbation with $V$ has been 
attempted to include two-magnon excitations by extending the UCL approximation,
\cite{Ament11} it may be logically inappropriate.

\subsection{Cluster model}

It is not easy to evaluate accurately $f_{0}^{(1)}(\omega_i)$ 
and $f_{m}^{(2)}(\omega_i)$'s.
In this paper, we use a cluster consisting of a central spin and of
8 neighboring spins, as shown in Fig.~\ref{fig.cluster}. 
The outer spins labeled as $\textbf{S}_{1}\sim \textbf{S}_{8}$ are assumed
under the staggered field from spins outside the cluster.
For the central spin ${\bf S}_{0}$ belonging to the A sublattice,
the initial-state Hamiltonian is given by
\begin{equation}
 H_{\rm mag} =
J\sum_{\langle i,j\rangle} {\bf S}_i\cdot{\bf S}_{j}
+ \frac{J}{2} \sum_{i=1}^{4} S_{i}^{z}
- 2 \frac{J}{2} \sum_{j=5}^{8} S_{j}^{z}, 
\label{eq.cluster-ham}
\end{equation}
which is represented by the matrix with $512\times 512$ dimensions.
The intermediate-state Hamiltonian is given just by eliminating ${\bf S}_0$
from Eq.~(\ref{eq.cluster-ham}), which is represented by the matrix with
$256\times 256$ dimensions.
Diagonalizing numerically the Hamiltonian matrices, we obtain the
eigenstates of $H_{\rm mag}$ and $H'$. 
We obtain, for example, 
$\langle S_{0}^{z}\rangle =0.394$ from Eq.~(\ref{eq.magnetization}). 
This value is an overestimate of the sublattice magnetization in comparison
with the $1/S$-expansion value in second order 0.307,\cite{Igarashi92-1}
probably due to the finite-size effect.
Using the eigenstates of $H_{\rm mag}$ and $H'$, we evaluate 
$f_{0}^{(1)}$ from Eqs.~(\ref{eq.B}) and (\ref{eq.BD}),
and $f_{m}^{(2)}$ from Eqs.~(\ref{eq.f21}) and (\ref{eq.f22}).
Since the cluster size is strongly subject to the boundary,
the values thus obtained may be considered
as a semi-quantitative estimate.

\begin{figure}
\includegraphics[width=8.0cm]{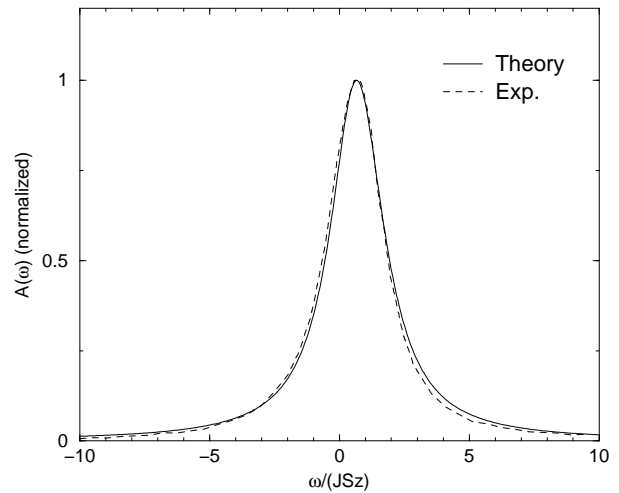}%
\caption{\label{fig.abs}
Absorption coefficient $A(\omega_i)$ as a function of photon energy $\omega_i$.
$\Gamma/(2J)=1.2$. The origin of energy is set to correspond to
$\omega_i=\epsilon_{\rm core}$. The dotted line represents the XAS 
experimental data at the Cu $L_{3}$ edge in Sr$_2$CuO$_2$Cl$_2$,
\cite{Guarise10}
where $J$ is assumed to be $130$ meV, and the experimental curve 
is shifted such that the peak position coincides with the calculation.
}
\end{figure}

Using the notation in the preceding subsection, we express
the $L_{2,3}$-absorption coefficient $A_j(\omega_i)$,
\begin{eqnarray}
 A_j(\omega_i) &=& \sum_{\alpha_i}P^{(0)}(j;\alpha_i,\alpha_i)
  \sum_{\sigma,\eta}|\langle \phi_{\eta}|\psi_0^{\sigma}\rangle|^2
  \nonumber \\
&\times& \frac{\Gamma/\pi}{(\omega_i+\epsilon_g-\epsilon_{\rm core}
  -\epsilon'_{\eta})^2+\Gamma^2}.
\label{eq.abs}
\end{eqnarray}
Substituting the eigenstates of the cluster model into Eq.~(\ref{eq.abs}),
we calculate $A_{j}(\omega_i)$.
Figure \ref{fig.abs} shows the calculated $A_{j}(\omega_i)$
as a function of photon energy. 
The origin of photon energy is $\omega_i=\epsilon_{\rm core}$. 
The dimensionless life-time broadening width of the core hole is
chosen as $\Gamma/(2J)=1.2$.
The calculated curve is found very close to the Lorentzian shape.
By comparison with the experimental curve for the Cu $L_{3}$-edge
in Sr$_2$CuO$_2$Cl$_2$, $\Gamma$ is estimated $\sim 2.4J$ with $J=130$ meV.

\subsection{Spin correlation function}

The core hole could be excited at all the Cu sites in the x-ray scattering 
event. For the contribution from the core-hole site ${\bf r}_{\ell}$,
we need to multiply weight $\exp(i{\bf q}\cdot{\bf r}_{\ell})$ 
(${\bf q}\equiv{\bf q}_i-{\bf q}_f$) to the amplitudes discussed in the
preceding subsections.
Collecting the contributions from each core-hole site in Eq.~(\ref{eq.optical}),
we finally obtain the expression of the RIXS spectra:
\begin{eqnarray}
 W(q_f\alpha_f;q_i\alpha_i) &=&
  \frac{w^4}{4\omega_i\omega_f} 
  \Bigl\{|P^{(1)}(j;\alpha_f,\alpha_i)|^2 Y^{(1)}(\omega_i;{\bf q},\omega)
\nonumber\\
   &+&|P^{(0)}(j;\alpha_f,\alpha_i)|^2 Y^{(2)}(\omega_i;{\bf q},\omega)
  \Bigr\},
\end{eqnarray}
where the spin-flip and the spin-conserving correlation functions are 
defined by 
\begin{eqnarray}
 Y^{(1)}(\omega_i;{\bf q},\omega)&=&\int_{-\infty}^{\infty}
 \langle Z^{(1)\dagger}(\omega_i;{\bf q},t)Z^{(1)}(\omega_i;{\bf q},0)
\rangle {\rm e}^{i\omega t}{\rm d}t, \nonumber \\
\\
 Y^{(2)}(\omega_i;{\bf q},\omega)&=&\int_{-\infty}^{\infty}
 \langle Z^{(2)\dagger}(\omega_i;{\bf q},t)Z^{(2)}(\omega_i;{\bf q},0)
\rangle {\rm e}^{i\omega t}{\rm d}t, \nonumber \\
\end{eqnarray}
where the angular bracket denotes the expectation value in the
ground state.
Operators $Z^{(1)}(\omega_i;{\bf q})$ and 
$Z^{(2)}(\omega_i;{\bf q})$ are given by
\begin{eqnarray}
 Z^{(1)}(\omega_i;{\bf q})&=& 2f_{0}^{(1)}(\omega_i)
      \bigl\{S_{a}^{x'}(-{\bf q})+S_{b}^{x'}(-{\bf q})\bigr\}
\label{eq.one} \\
 Z^{(2)}(\omega_i;{\bf q})
    &=& f_{2}^{(2)}(\omega_i)\Bigl\{
    (X^{z'}S^{z'})_{a}(-{\bf q})+(X^{z'}S^{z'})_{b}(-{\bf q})
\nonumber\\
    &+& (X^{x'}S^{x'})_{a}(-{\bf q})+(X^{x'}S^{x'})_{b}(-{\bf q}) 
\nonumber\\
    &+& (X^{y'}S^{y'})_{a}(-{\bf q})+(X^{y'}S^{y'})_{b}(-{\bf q})
    \Bigr\}.
\label{eq.two} 
\end{eqnarray}
The Fourier transforms of $S_{0}^{x'}$ and
$X^{x'(y',z')}S_0^{x'(y',z')}$ are defined separately 
for the A and B sublattices
and discriminated by subscripts $a$ and $b$, respectively.
For example, by multiplying weight 
${\rm e}^{i{\bf q}\cdot{\bf r}_{\ell}}$, we have
\begin{eqnarray}
 (X^{x'}S^{x'})_{a}(-{\bf q}) &=&\sum_{i\in{\rm A}}
 \frac{1}{2}\sum_{\delta}S_{i+\delta}^{x'}S_{i}^{x'}
 {\rm e}^{i{\bf q}\cdot {\bf r}_i}, \\
 (X^{x'}S^{x'})_{b}(-{\bf q}) &=& \sum_{j\in{\rm B}}
 \frac{1}{2}\sum_{\delta}S_{j+\delta}^{x'}S_{j}^{x'}
 {\rm e}^{i{\bf q} \cdot {\bf r}_j}.
\end{eqnarray}
Note that $Y^{(1)}(\omega_i;{\bf q},\omega)$ is proportional to
the dynamical structure factor of the transverse spin, while
$Y^{(2)}(\omega_i;{\bf q},\omega)$ is proportional to the ``exchange"-type
spin-spin correlation function discussed in the magnetic excitations 
in the $K$-edge RIXS.\cite{Nagao07,Ament07}

\section{\label{sect.4}Spin correlation function within the $1/S$ expansion}

Spin-flip and spin-conserving excitations derived in the preceding section 
could freely propagate in the crystal in the final state because of 
the absence of core hole.
Therefore, the cluster model with small size would not work well 
in the final state. We exploit the $1/S$-expansion method
in the study of the final state. Notations and several relations required
in the present study are briefly summarized in Appendix.

\begin{figure}[t]
\includegraphics[width=8.0cm]{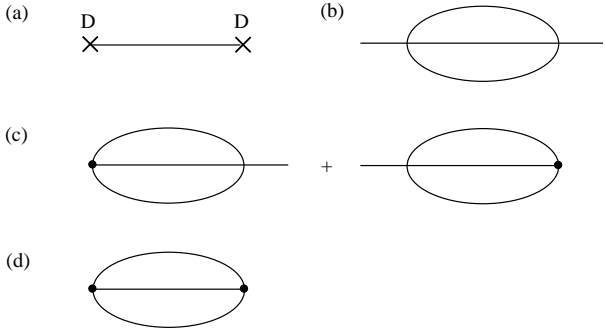}\\
\caption{\label{fig.diagram}
Diagrams for the time-ordered Green function 
$\tilde{Y}^{(1)}(\omega_i;{\bf q},\omega)$ required to evaluate the correction
up to the $1/(2S)^2$. Solid lines represent the unperturbed Green
functions $G_{\alpha\alpha}^{(0)}({\bf k},\omega)$ or
$G_{\beta\beta}^{(0)}({\bf k},\omega)$, 
on which the arrows and the type ($\alpha$ or $\beta$) are omitted. 
The crosses in diagram (a) represent the spin reduction factor.
Solid circles represent the three-magnon terms given
by Eq.~(\ref{eq.z1}) and their Hermitian conjugates.
}
\end{figure}

\subsection{Spin-flip excitation spectra} 

We express the spin-flip operator $Z^{(1)}(\omega_i;{\bf q})$ 
in terms of magnon operators by using Eqs.~(\ref{eq.boson1})-(\ref{eq.magnon}).
Since $Z^{(1)}(\omega_i;{\bf q})$ is proportional to 
$S_a^{x'}(-{\bf q})+S_b^{x'}(-{\bf q})$,
the derivation is parallel to those of Eqs.~(6.9) and (6.10) in Ref. 
\onlinecite{Igarashi05} in the study of the dynamical structure factor.
In the following, we simply write down the result. 

For ${\bf q}$ being inside the first magnetic Brillouin zone (MBZ), 
we have
\begin{widetext}
\begin{eqnarray}
 Z^{(1)}(\omega_i;{\bf q}) &=& \sqrt{\frac{N}{2}}\sqrt{2S}
 \frac{1}{2}f_{0}^{(1)}(\omega_i)(\ell_{\bf q}+m_{\bf q})
\left[ D(\alpha_{\bf -q}^{\dagger}+\beta_{\bf -q}^{\dagger})
           +(\alpha_{\bf q}+\beta_{\bf q})
\right. 
   \nonumber\\
  & &
-\frac{1}{2S}\frac{2}{N}\sum_{234}\delta_{\bf G}({\bf q}+2-3-4)
     \frac{1}{2}\ell_{\bf q}\ell_{2}\ell_{3}\ell_{4}
\nonumber\\
& & \left. \times
 \left\{M_{{\bf q}234}^{(1)}+{\rm sgn}(\gamma_{\bf G})
   M_{{\bf q}234}^{(2)}\right\} 
 \left\{\beta_{2}^{\dagger}\alpha_{-3}^{\dagger}\alpha_{-4}^{\dagger}
+ \beta_{-2}\alpha_{3}\alpha_{4}
         + {\rm sgn}(\gamma_{\bf G}) \left( 
    \alpha_{2}^{\dagger}\beta_{-3}^{\dagger}\beta_{-4}^{\dagger}
       +  \alpha_{-2}\beta_{3}\beta_{4} \right) \right\} \right],
\label{eq.z1}
\end{eqnarray}
\end{widetext}
where
\begin{eqnarray}
 M^{(1)}_{{\bf q}234} &=& -x_2+ {\rm sgn}(\gamma_{\bf G})x_{\bf q}x_3x_4, \\
 M^{(2)}_{{\bf q}234} &=& x_3x_4 - {\rm sgn}(\gamma_{\bf G})x_{\bf q}x_2.
\end{eqnarray}
The Kronecker delta $\delta_{\textbf{G}}(\textbf{q}+2-3-4)$ 
indicates the conservation of momenta within a reciprocal lattice vector 
$\textbf{G}$ and  ${\rm sgn}(\gamma_{\bf G})$ denotes the sign 
of $\gamma_{\bf G}$. $x_{\bf k}$ is defined by Eq.~(\ref{eq.xk}).
The \emph{spin reduction factor} $D$, which is related to the zero-point 
reduction of spin, is given by
\begin{equation}
 D=1-\frac{\Delta S}{2S}-\frac{1}{4}\frac{\Delta S(1+3\Delta S)}{(2S)^2},
 \label{eq.D}
\end{equation}
with
\begin{equation}
 \Delta S = \frac{1}{N}\sum_{\bf k}(\epsilon_{\bf k}^{-1}-1).
\end{equation}
For the square lattice, $\Delta S=0.1966$.

Introducing the time-ordered Green function defined by
$\tilde{Y}^{(1)}(\omega_i;{\bf q},\omega)
\equiv -i\int\langle T(Z^{(1)\dagger}(\omega_i;{\bf q},t)
Z^{(1)}(\omega_i;{\bf q},0)\rangle{\rm e}^{i\omega t}{\rm d}t$,
we expand the Green function up to the second order with $1/S$,
which diagrams are shown in Fig.~\ref{fig.diagram}.
We obtain the correlation function from the time-ordered Green function
by the fluctuation-dissipation theorem, 
$Y^{(1)}(\omega_i;{\bf q},\omega)=-2{\rm Im}
\tilde{Y}^{(1)}(\omega_i;{\bf q},\omega)$. Here, $\textrm{Im} X$ means
imaginary part of quantity $X$.
The correlation function 
is found to consist of the $\delta$-function peak 
of the one-magnon excitation and the continuum of three-magnon excitations,
that is, 

\begin{figure}[t]
\includegraphics[width=7.50cm]{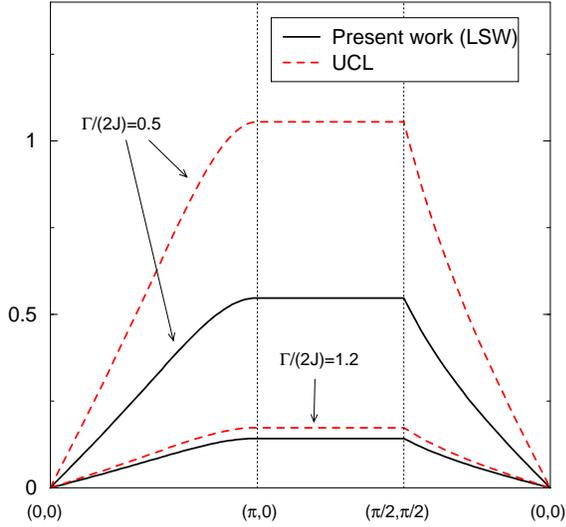}
\caption{\label{fig.one-mag.LSW}
(Color online) 
One-magnon intensity $y_{1}(\omega_i;{\bf q})/[N(2S)]$ within the LSW contribution,
as a function of ${\bf q}$ along symmetry lines,
in comparison with the results of the UCL approximation 
(red broken line).
The $\omega_i$ is set to give rise to the peak in the absorption spectra.
}
\end{figure}

\begin{equation}
Y^{(1)}(\omega_i;{\bf q},\omega)=
 y_{1}(\omega_i;{\bf q})(2\pi)\delta(\omega-\epsilon_{\bf q})
+y_3(\omega_i;{\bf q},\omega).
\label{eq.one-mag.nonlinear}
\end{equation}
Within the second-order of $1/S$, the one-magnon spectral weight 
$y_1({\bf q})$ may be expressed as 
\begin{eqnarray}
 y_1(\omega_i;{\bf q}) &=& N(2S)
  |f_1^{(1)}(\omega_i)m_{\bf q}+f_2^{(1)}(\omega_i)\ell_{\bf q}|^2 \nonumber \\
&\times&
 \left\{1+\frac{1}{2S}y_{1,1}({\bf q})
              +\frac{1}{(2S)^2}y_{1,2}({\bf q}) \right\},
\label{eq.y1}
\end{eqnarray}
where
\begin{widetext}
\begin{eqnarray}
 y_{1,1}({\bf q})&=& \left(-2\Delta S\right),\label{eq.one-mag.first}\\
 y_{1,2}({\bf q})&=& -\frac{1}{2}\Delta S (1+\Delta S) 
 + \frac{-1}{\epsilon_{\bf q}}
      \Sigma_{\alpha\beta}^{(2)}({\bf q},\epsilon_{\bf q}) \nonumber\\
 &+& \left(\frac{2}{N}\right)^2\sum_{\bf pp'}
     2\ell_{\bf q}^2\ell_{\bf p}^2\ell_{\bf p'}^2\ell_{\bf q+p-p'}^2
\left[\frac{-|B^{(4)}_{\bf q,p,p',[q+p-p']}|^2}
{(\epsilon_{\bf q}-\epsilon_{\bf p}-\epsilon_{\bf p'}-\epsilon_{\bf q+p-p'})^2}
  +\frac{|B^{(6)}_{\bf q,p,p',[q+p-p']}|^2}
{(\epsilon_{\bf q}+\epsilon_{\bf p}+\epsilon_{\bf p'}+\epsilon_{\bf q+p-p'})^2}
 \right] \nonumber\\
  &+& \left(\frac{2}{N}\right)^2\sum_{\bf pp'}
    2\ell_{\bf q}^2\ell_{\bf p}^2\ell_{\bf p'}^2\ell_{\bf k+p-p'}^2
\left[\frac{B^{(4)}_{\bf q,p,p',[q+p-q']}N^{(1)}_{\bf q,p,p',[q+p-p']}}
   {(\epsilon_{\bf q}-\epsilon_{\bf p}-\epsilon_{\bf p'}-\epsilon_{\bf q+p-p'})}
   -\frac{{\rm sgn}(\gamma_{\bf G})B^{(6)}_{\bf q,p,p',[q+p-p']}
           N^{(2)}_{\bf q,p,p',[q+p-p']}}
   {\epsilon_{\bf q}+\epsilon_{\bf p}+\epsilon_{\bf p'}+\epsilon_{\bf q+p-p'}}
  \right]. \nonumber \\
\label{eq.one-mag.second}
\end{eqnarray}
with 
\begin{eqnarray}
  N^{(1)}_{\bf q,p,p'[q+p-p']} &=& 
   M_{\bf q,p,p',[q+p-p']}^{(1)}
  +  {\rm sgn}(\gamma_{\bf G})M_{\bf q,p,p',[q+p-p']}^{(2)}, \\
  N^{(2)}_{\bf q,p,p'[q+p-p']} &=&  
    M_{\bf q,p,p',[q+p-p']}^{(1)} +
    {\rm sgn}(\gamma_{\bf G})M_{\bf q,p,p',[q+p-p']}^{(2)}.
\end{eqnarray}
\end{widetext}
Here ${\bf [q+p-p']}$ stands for ${\bf q+p-p'}$ reduced to the first MBZ by a
reciprocal vector ${\bf G}$, that is, ${\bf [q+p-p']}={\bf q+p-p'-G}$.
Equation (\ref{eq.one-mag.first}) and the first term of 
Eq.~(\ref{eq.one-mag.second}) arise from the spin-reduction factor
with the diagram (a). The second and the third terms of 
Eq.~(\ref{eq.one-mag.second}) arise from the diagram (b), and the fourth term
arises from the diagram (c). No contribution comes out from the diagram (d).
A systematic study of the $1/S$ expansion for the dynamical structure factor
has been carried out in Ref. \onlinecite{Igarashi05}.

Figure \ref{fig.one-mag.LSW} shows the one-magnon intensity 
$y_{1}(\omega_i;{\bf q})$ within the LSW contribution,
as a function of ${\bf q}$ along symmetry lines,
where $\omega_i$ is set to give rise to the peak in the absorption spectra.
Since it is proportional to the dynamical structure factor,
its ${\bf q}$-dependence is well known.
The one-magnon intensity vanishes with ${\bf q}\to 0$, while diverges
with ${\bf q}\to (\pi,\pi)$. 
The difference from the result of the UCL approximation is only its magnitude;
the difference becomes larger with decreasing values of $\Gamma$.

\begin{figure}[t]
\includegraphics[width=8.0cm]{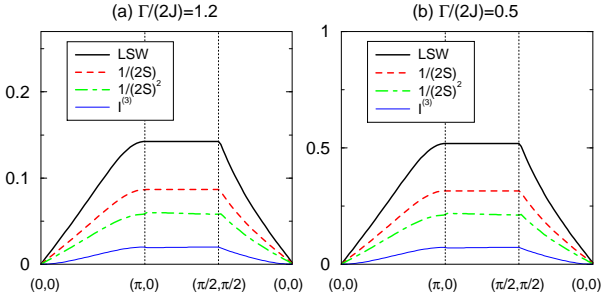}
\caption{\label{fig.one-mag.int}
(Color online) One-magnon intensity $y_{1}(\omega_i;{\bf q})/[N(2S)]$
up to the first 
(red broken line) and the second (green dot-dashed line) 
order of $1/S$, as a function of ${\bf q}$ along symmetry lines.
The $\omega_i$ is set to give rise to the peak in the absorption spectra.
The integrated intensity of three-magnon continuum $I^{(3)}(\omega_i;{\bf q})$
is also shown by blue thin line.
}
\end{figure}

Figure \ref{fig.one-mag.int} shows the one-magnon intensity evaluated up to the
first and the second order of $1/S$ [Eqs.~(\ref{eq.one-mag.first})
and (\ref{eq.one-mag.second})].
The intensities are found strongly reduced from the zeroth-order values.
The $1/S$ expansion has already been carried out to the transverse
component of the dynamical structure factor
(see Fig.~5 in Ref. \onlinecite{Igarashi05}).

Now we discuss the continuum spectra of three-magnon excitations.
Analyzing carefully the diagrams in Fig.~\ref{fig.diagram}, 
we finally obtain from the second-order contribution with respect to $1/S$ as
\begin{widetext}
\begin{eqnarray}
 y_3(\omega_i;{\bf q},\omega) &=& N(2S)
  |f_1^{(1)}(\omega_i)m_{\bf q}+f_2^{(1)}(\omega_i)\ell_{\bf q}|^2 
\frac{1}{(2S)^2}\left(\frac{2}{N}\right)^2\sum_{\bf p,p'}
  2\pi\delta(\omega-\epsilon_{\bf p}-\epsilon_{\bf p'}-\epsilon_{\bf q+p-p'}) 
\nonumber\\
 &\times& \frac{1}{2}\ell_{\bf q}^2\ell_{\bf p}^2\ell_{\bf p'}^2\ell_{\bf q+p-p'}^2
\Biggl| M_{\bf q,p,p',[q+p-p']}^{(1)}+
   {\rm sgn}(\gamma_{\bf G})M_{\bf q,p,p',[q+p-p']}^{(2)} \nonumber\\
 &-&2\left\{ \frac{B_{\bf q,p,p',[q+p-p']}^{(4)}}
   {\epsilon_{\bf q}-\epsilon_{\bf p}-\epsilon_{\bf p'}-\epsilon_{\bf q+p-p']}}
 + \frac{{\rm sgn}(\gamma_{\bf G})
      B_{\bf q,p,p',[q+p-p']}^{(6)}}
   {\epsilon_{\bf q}+\epsilon_{\bf p}+\epsilon_{\bf p'}+\epsilon_{\bf q+p-p'}}
   \right\}\Biggr|^2.
\label{eq.y3}
\end{eqnarray}
\end{widetext}
Note that Eq.~(\ref{eq.y3}) is, except for
the prefactor, equal to Eq.~(6.19) in Ref.~\onlinecite{Igarashi05}, 
as it should be.
The total intensity of three-magnon excitations is given by
\begin{equation}
 I^{(3)}(\omega_i;{\bf q})=\int_{0}^{\infty}
  y_3(\omega_i;{\bf q},\omega)\frac{{\rm d}\omega}{2\pi}.
\end{equation}
Figures \ref{fig.one-mag.int} (a) and (b) show
calculated $I^{(3)}(\omega_i;{\bf q})$ 
along symmetry lines for ${\bf q}$ with $\omega_i$ corresponding to the peak in the absorption
spectra. It is found that its intensity is about 
20\%-30\% of the one-magnon intensity.
 
\begin{figure}
\includegraphics[width=8.0cm]{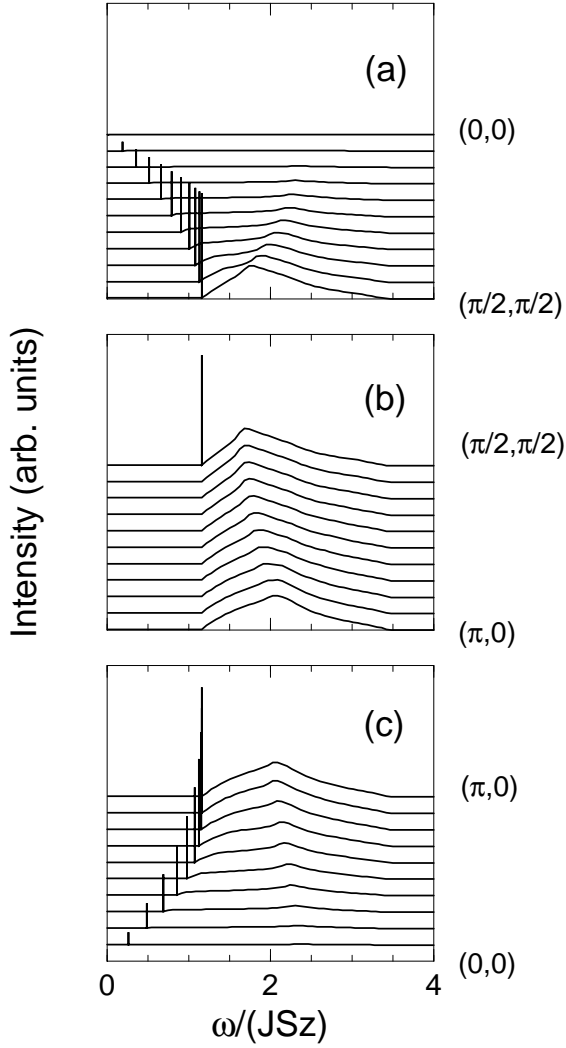}%
\caption{\label{fig.one-mag.disp}
Spin-flip correlation function 
$Y^{(1)}(\omega_i;{\bf q},\omega)/[N(2S)]$
as a function of energy loss $\omega$ along symmetry lines for ${\bf q}$.
Vertical solid lines represent the $\delta$-function
peak of the one-magnon excitation. On panel (b), only one vertical line is 
shown, since the peak positions and the weights of the one-magnon excitations
are nearly the same for $(\pi,0)-(\pi/2,\pi/2)$.
}
\end{figure}

The magnon energy $\epsilon_{\bf q}$ is corrected as
$\tilde{\epsilon}_{\bf q}=(1+A/2S)\epsilon_{\bf q}$
within the first-order in $1/S$. 
The second-order correction in $1/S$, which is known to be rather small
(see Ref.~\onlinecite{Igarashi92-1} for the details).
We replace $\delta(\omega-\epsilon_{\bf q})$ by
$\delta(\omega-\tilde{\epsilon}_{\bf q})$
in Eq.~(\ref{eq.one-mag.nonlinear}) and 
$\delta(\omega-\epsilon_{\bf p}-\epsilon_{\bf p'}-\epsilon_{q+p-p'})$
by
$\delta(\omega-\tilde{\epsilon}_{\bf p}-\tilde{\epsilon}_{\bf p'}-\tilde{\epsilon}_{q+p-p'})$
in Eq.~(\ref{eq.y3}), respectively. 
Figure \ref{fig.one-mag.disp} shows the spin-flip correlation function
$Y^{(1)}(\omega_i;{\bf q},\omega)/N(2S)$ as a function of $\omega$
along symmetry lines for ${\bf q}$.
The spectra consist of the $\delta$-function peak shown by vertical lines 
and the continuum of three-magnon excitations.

\subsection{Spin-conserving excitation spectra}

The two-magnon operator $Z^{(2)}(\omega_i;{\bf q})$ defined by 
Eq.~(\ref{eq.two}) is rewritten in terms of magnon operators
by using Eqs.~(\ref{eq.boson1})-(\ref{eq.magnon}).
The result is summarized as
\begin{equation}
 Z^{(2)}(\omega_i;{\bf q}) = (2S)\sum_{\bf k} N(\omega_i;{\bf q},{\bf k})
  \alpha_{[{\bf q+k}]}^{\dagger}\beta_{\bf -k}^{\dagger} + \cdots,
\label{eq.z2}
\end{equation}
with
\begin{widetext}
\begin{eqnarray}
 N(\omega_i;{\bf q},{\bf k}) 
 &=& f_{2}^{(2)}(\omega_i)\Bigl\{
 (1+\gamma_{\bf q})(\ell_{[{\bf q+k}]}m_{\bf k}
  +{\rm sgn}(\gamma_{\bf G}) \ell_{\bf k} m_{[{\bf q+k}]}) \nonumber\\
 &+& \ell_{[{\bf q+k}]}\ell_{\bf k}
  (\gamma_{\bf k}+{\rm sgn}(\gamma_{\bf G})\gamma_{[{\bf q+k}]}) 
 + m_{[{\bf q+k}]}m_{\bf k}
  (\gamma_{[{\bf q+k}]}+{\rm sgn}(\gamma_{\bf G})\gamma_{\bf k})
  \Bigr\},
\end{eqnarray}
\end{widetext}
where $[\textbf{q}+\textbf{k}]$ stands for the $\textbf{q}+\textbf{k}$ 
reduced in the first MBZ by a reciprocal lattice vector $\textbf{G}$, that is,
$[\textbf{q}+\textbf{k}]=\textbf{q}+\textbf{k}-\textbf{G}$.
In deriving Eq.~(\ref{eq.z2}), the non-linear terms in the expansion of
$S^{\pm}$ with magnon operators have been neglected, since the consistent
analysis within the $1/S$ expansion is quite complicated. Such terms
may cause the intensity transfer to the four-magnon excitations.
This type of correlation function has been studied for the magnetic
excitations in the $K$-edge RIXS.\cite{Nagao07,Ament07}

\begin{figure}
\includegraphics[width=8.0cm]{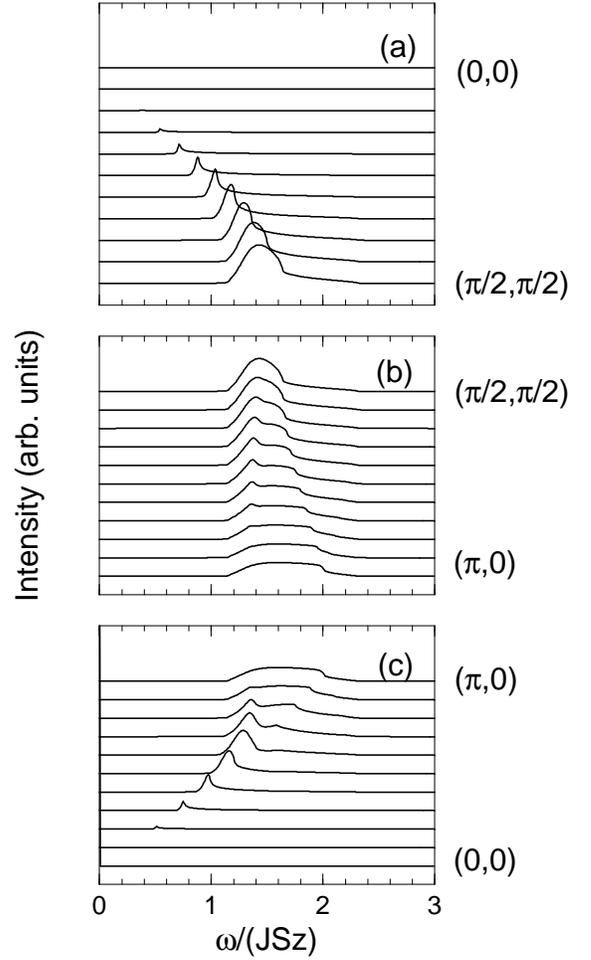}%
\caption{\label{fig.two-mag.disp}
Spin-conserving correlation function 
$Y^{(2)}(\omega_i;{\bf q},\omega)/[N(2S)^2]$
as a function of energy loss $\omega$ along symmetry lines for ${\bf q}$.
}
\end{figure}

Since two-magnons are excited closely to each other around the core-hole site,
the magnon-magnon interaction would be important.
Introducing the two-magnon Green function,
\begin{equation}
F(\textbf{q},\omega;\textbf{k},\textbf{k}')
=-i \int \textrm{e}^{i \omega t} {\rm d}t
\langle T[\beta_{-\textbf{k}}(t)
          \alpha_{[\textbf{q}+\textbf{k}]}(t)
          \alpha_{[\textbf{q}+\textbf{k}']}^{\dagger}
          \beta_{-\textbf{k}'}^{\dagger}] \rangle, 
\end{equation}
we take account of scattering of two magnons through the term 
$B_{1234}^{(3)}$ [Eq. (\ref{eq.A16})].
Applying the addition theorem of trigonometric functions to factors
such as $\gamma_{2-4}$ in $B^{(3)}_{1234}$, we could transform 
$B^{(3)}_{1234}$ into a separable form. Thereby we obtain the t-matrix 
by summing up the ladder diagrams in a closed form.
We have already explained this procedure in the analysis of the two-magnon
spectra in the $K$-edge RIXS. See Ref.~\onlinecite{Nagao07} for the details.
Once we obtain the Green function, by the fluctuation-dissipation 
theorem, we obtain the correlation function as
\begin{eqnarray}
Y^{(2)}(\omega_i;\textbf{q},\omega) 
&=& (2S)^2 \sum_{\textbf{k}} \sum_{\textbf{k}'}
N^{*}(\omega_i;\textbf{q},\textbf{k}) N(\omega_i;\textbf{q},\textbf{k}')
\nonumber \\
&\times&
(-2){\rm Im}F(\textbf{q},\omega;\textbf{k},\textbf{k}').
\label{eq.yt}
\end{eqnarray}
Figure \ref{fig.two-mag.disp} shows $Y^{(2)}(\omega_i;{\bf q},\omega)$ 
as a function of $\omega$ along symmetry lines for ${\bf q}$
with $\omega_i$ corresponding to the peak in the absorption spectra.
 
\begin{figure}
\includegraphics[width=8.0cm]{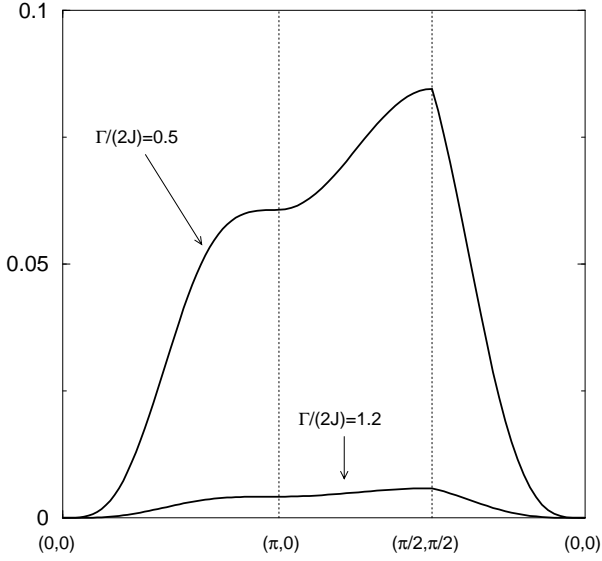}%
\caption{\label{fig.two-mag.int}
Frequency-integrated correlation function 
$Y^{(2)}(\omega_i;{\bf q})/[N(2S)^2]$
as a function of ${\bf q}$ along symmetry lines.
The $\omega_i$ is set to give rise to the peak in the absorption spectra.
}
\end{figure}

The frequency-integrated correlation function is not changed 
by the presence of the magnon-magnon interaction.
Neglecting the interaction in the calculation of 
$F(\textbf{q},\omega;\textbf{k},\textbf{k}')$ in Eq.~(\ref{eq.yt}),
we simply obtain
\begin{eqnarray}
Y^{(2)}(\omega_i;{\bf q})&\equiv& \int Y^{(2)}(\omega_i:{\bf q},\omega)
                                \frac{{\rm d}\omega}{2\pi}
	\nonumber \\
&=& (2S)^2\sum_{\bf k}|N(\omega_i;{\bf q},{\bf k})|^2.
\end{eqnarray}
Figure \ref{fig.two-mag.int} shows $Y^{(2)}(\omega_i;{\bf q})/N(2S)^2$ 
as a function of ${\bf q}$ along symmetry lines with $\omega_i$ 
corresponding to the peak in the absorption spectra.
The values are about one order of magnitude smaller than those of 
$Y^{(1)}(\omega_i;{\bf q})/[N(2S)]$.

\section{\label{sect.5} Spectral shape in comparison with experiments}

In this section, we analyze specifically the Cu $L_{3}$ edge spectra 
in Sr$_2$CuO$_2$Cl$_2$.
According to the experimental setup shown in Fig.~1(a) 
of Ref.~\onlinecite{Guarise10},
the photon comes onto the $a$-$b$ ($x$-$y$) plane,
and is scattered with the angle $130$ degrees to the incident direction.
The $a$-$c$ ($x$-$z$) plane is set as the scattering plane.
The polarization vector of the incident photon is then expressed as 
$\mbox{\boldmath{$\alpha$}}_i=(0,-1,0)$ for the $\sigma$ polarization and
$\mbox{\boldmath{$\alpha$}}_i=(\chi_i^{\pi},0,\zeta_i^{\pi})$ 
for the $\pi$ polarization.
Similarly,
the polarization of the scattered photon is expressed as
$\mbox{\boldmath{$\alpha$}}_f=(0,-1,0)$ for the $\sigma'$ polarization 
and
$\mbox{\boldmath{$\alpha$}}_f=(\chi_f^{\pi},0,\zeta_f^{\pi})$ for the $\pi'$ polarization.
Thereby we have from Table \ref{table.2}, 
\begin{widetext}
\begin{eqnarray}
P^{(1)}(3/2,\alpha^f,\alpha^i)&=& 0, \quad
P^{(0)}(3/2,\alpha^f,\alpha^i)=\frac{2}{15}, \quad (\sigma\to\sigma'),
\label{eq.pol1}\\
P^{(1)}(3/2,\alpha^f,\alpha^i)&=&-\frac{i}{15}\chi_f^{\pi}, \quad
P^{(0)}(3/2,\alpha^f,\alpha^i)=0, \quad (\sigma\to\pi'), \\
P^{(1)}(3/2,\alpha^f,\alpha^i)&=&0, \quad
P^{(0)}(3/2,\alpha^f,\alpha^i)=\frac{2}{15}\chi_f^{\pi}\chi_{i}^{\pi},
\quad (\pi\to\pi'), \\
P^{(1)}(3/2,\alpha^f,\alpha^i)&=&\frac{i}{15}\chi_i^{\pi}, \quad
P^{(0)}(3/2,\alpha^f,\alpha^i)=0, \quad (\pi\to\sigma').
\label{eq.pol4}
\end{eqnarray}
The polarization is separated with the incident photon, but not separated 
with the scattered photon in the experiment. 
In this situation, the RIXS spectra, which depend on the polarization of 
the incident photon, are expressed from Eqs.~(\ref{eq.pol1})-(\ref{eq.pol4})
as
\begin{equation}
I(\omega_i;{\bf q},\omega)=\frac{w^4}{4\omega_i\omega_f} \times 
\left\{ \begin{array}{ll}
 \left[\left(\frac{\chi_f^{\pi}}{15}\right)^2
Y^{(1)}(\omega_i;{\bf q},\omega)
     +\left(\frac{2}{15}\right)^2
Y^{(2)}(\omega_i;{\bf q},\omega)\right],
& (\sigma-{\rm pol.}), \\
 \left[\left(\frac{\chi_i^{\pi}}{15}\right)^2
Y^{(1)}(\omega_i;{\bf q},\omega)
     +\left(\frac{2}{15}\right)^2\left(\chi_f^{\pi}\chi_i^{\pi}\right)^2
  Y^{(2)}(\omega_i;{\bf q},\omega)\right],
&  (\pi-{\rm pol.}), \\
\end{array} \right. ,
\end{equation}
\end{widetext}
where ${\bf q}$ is regarded as the component (of the transferred momentum)
projected onto the $a$-$b$ plane.
For the spin-flip excitations, the intensity in the $\pi$ polarization
is larger (or smaller) by a factor $(\chi_i^{\pi}/\chi_f^{\pi})^2$ than
that in the $\sigma$ polarization. For the spin-conserving excitations, 
the intensity in the $\pi$ polarization is smaller by a factor
$(\chi_f^{\pi}\chi_i^{\pi})^2$ than that in the $\sigma$ polarization.
Notice that, in both cases, 
the intensity ratio between in the $\pi$ polarization
and in the $\sigma$ polarization is completely determined from the 
scattering geometry.  That is, the ratio is independent of the values of 
the correlation functions.

\begin{figure}
\includegraphics[width=8.0cm]{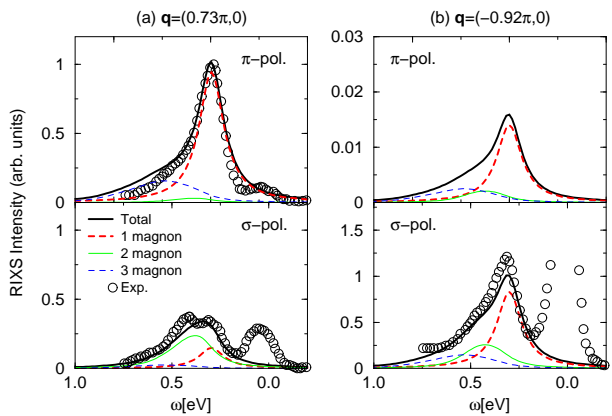}%
\caption{\label{fig.shape73}
(Color online)
RIXS spectra as a function of energy loss $\omega$ for the momentum
transfer projected onto the $a-b$ plane: panel (a) for 
${\bf q}=(0.73\pi,0)$, and panel (b) for ${\bf q}=(-0.92\pi,0)$.
The incident photon energy $\omega_i$ is set
to give rise to the peak in the absorption spectra.
$J=130$ meV, and the calculated spectra are convoluted with the Lorentzian
function with the half width of half maximum $78$ meV.
The experimental data for Sr$_2$CuO$_2$Cl$_2$ are taken from
Ref.~\onlinecite{Guarise10}; the peak height is 
adjusted to coincide with the calculated one in the $\pi$ polarization
for ${\bf q}=(0.73\pi,0)$,  while it is adjusted with 
the calculated one in the $\sigma$ polarization for 
${\bf q}=(-0.92\pi,0)$.
The thick solid (black) lines represent the total intensities.
The thick dotted (red), thin solid (green) and 
thin dotted (blue) lines
correspond to the one, two, and three magnon contributions,
respectively.}
\end{figure}

Figure \ref{fig.shape73} shows the calculated result 
with $\omega_i$ giving rise to the peak in the absorption spectra.
For comparison with the experiment,\cite{Guarise10} 
the exchange coupling constant $J$ is assumed to be $130$ meV, 
which is nearly the same as the one estimated by the inelastic neutron
scattering experiment in La$_2$CuO$_4$.\cite{Aeppli89}
The calculated spectra are convoluted with the Lorentzian function
with the half width of half maximum $78$ meV in accordance 
with the experimental resolution.
Panel (a) shows the spectra for ${\bf q}=(0.73\pi,0)$.
According to the experimental setup, we have $\chi_i^{\pi}=-0.95$ and
$\chi_f^{\pi}=0.38$.
The experimental data are drawn such that the peak height 
coincides with the calculated peak in the $\pi$ polarization.
In the $\pi$ polarization, the spin-conserving contribution is 
suppressed by the relative weight $(2\chi_f^{\pi})^2 = 0.58$, and
the spectra are dominated by the spin-flip contribution.
The spin-flip contribution includes the three-magnon continuum with
the energy higher than the single-magnon peak.
This makes the spectral shape asymmetric in agreement with the 
experimental data in the $\pi$ polarization.  
In the $\sigma$ polarization, the spin-flip contribution becomes smaller by 
a factor $(\chi_f^{\pi}/\chi_i^{\pi})^2=0.16$ than that in the $\pi$
polarization, while the spin-conserving contribution becomes larger by a factor
$1/(\chi_f^{\pi}\chi_i^{\pi})^2=7.7$ than that in the $\pi$ polarization,
and thereby the two-magnon intensity becomes larger in comparison with
the one-magnon intensity. 
In the experiment, the spectra consist of two peaks;\cite{Guarise10}
the low-energy peak is considered to come from the one-magnon excitation,
which intensity is estimated  to be smaller by a factor $0.32$ than 
the one-magnon intensity in the $\pi$ polarization.
This value is about twice the theoretical value, and
the reason for this discrepancy is not known, since the value is determined 
by the geometry. 
In the calculated spectra, the one-magnon and two-magnon peaks are closely
located, forming a single peak. However, the spectral shape is asymmetric 
with a broad width in agreement with the experiment, and the total intensity 
is also in good agreement with the experiment.

Panel (b) shows the spectra for ${\bf q}=(-0.92\pi,0)$.
According to the experimental setup, we have
$\chi_i^{\pi}=-0.11$, $\chi_f^{\pi}=0.85$.
Since $(\chi_i^{\pi}/\chi_f^{\pi})^2=0.016$ and 
$(\chi_i^{\pi}\chi_f^{\pi})^2=0.0087$,
the intensity in the $\pi$ polarization is two order of magnitude smaller
than that in the $\sigma$ polarization.
In the $\sigma$ polarization, the two-magnon intensity has a considerable
weight, resulting in a shoulder in the high energy side.
The overall shape with a considerable width agrees well with the experiment.

\section{\label{sect.6} Concluding Remarks}

We have studied the magnetic excitations in the $L$-edge RIXS in undoped
cuprates. 
We have analyzed in detail the second-order dipole allowed process 
with paying attention to the strong perturbation through the 
intermediate state, in which there is no spin degree of freedom 
at the core-hole site.
In this situation, it is not logically appropriate to make a perturbation
calculation with the terms involving the spin degree of freedom lost
in the intermediate state. 
Within the approximation that the perturbation due to
the intermediate state is not extending to neighboring sites,
we derive the spin-flip final state expressed as
$\mbox{\boldmath{$\alpha$}}_{f\perp}\times
\mbox{\boldmath{$\alpha$}}_{i\perp}\cdot{\bf S}_0|g\rangle$ 
in the scattering channel with changing the polarization,
which leads to the RIXS spectra expressed as 
the dynamical structure factor of the transverse spin component.
In the scattering channel without changing the polarization,
we have assumed a spherical form of the spin-conserving final state, 
${\bf X}\cdot{\bf S}_0|g\rangle$, which leads to the RIXS spectra
expressed as the `exchange'-type multi-spin correlation function.
We have numerically evaluated the transition amplitudes for both 
the spin-flip and the spin-conserving final 
states on a finite-size cluster centered at the core-hole site.

Since no core hole exists in the final state,
the spin excitations could move around the crystal.
We have treated the itinerant spin excitations by means of the $1/S$-expansion
method, which is known to work for treating the quantum fluctuation in the 
two-dimensional Heisenberg antiferromagnet.\cite{Igarashi92-1,Igarashi05}
For the spin-flip excitations, having expanded the spin-flip operators
up to the second order of $1/S$, we have obtained the three-magnon excitations
in addition to the one-magnon excitations.
This gives rise to a considerable reduction of the one-magnon intensity 
as well as the intensity transfer to the three-magnon continuum.
For the spin-conserving excitations, we have taken into account the
interaction between magnons by summing up the ladder diagrams.

We have analyzed the Cu $L_3$-edge spectra in Sr$_2$CuO$_2$Cl$_2$
on the basis of these results. The two- and three-magnon excitations
give rise to substantial intensities in the high energy side of
the one-magnon peak as a function of energy loss,
in good agreement with the experiment.\cite{Guarise10}
We hope that the similar analyses are applied to the RIXS spectra in other 
materials and clarify the nature of magnetic excitations in future.

\begin{acknowledgments}
The authors thank to Prof. J. van den Brink for useful discussion.
This work was partially supported by a Grant-in-Aid for Scientific Research 
from the Ministry of Education, Culture, Sports, Science and Technology
of the Japanese Government.

\end{acknowledgments}

\appendix
\section{1/S expansion}
We summarize briefly the $1/S$ expansion method in the Heisenberg 
antiferromagnet. For details, see Refs.~\onlinecite{Igarashi92-1}
and \onlinecite{Igarashi05}. The $x$, $y$, $z$ axes below are interpreted
as the $x'$, $y'$, $z'$ axes in the text.

\subsubsection{Hamiltonian}
Assuming two sublattices in the antiferromagnetic ground state, 
we express spin operators by boson operators as
\begin{eqnarray}
 S_i^z &=& S - a_i^\dagger a_i ,  
 \label{eq.boson1}\\
 S_i^+ &=& (S_i^-)^\dagger = \sqrt{2S}f_i(S)a_i , \\
 S_j^z &=& -S + b_j^\dagger b_j ,\\
 S_j^+ &=& (S_j^-)^\dagger = \sqrt{2S}b_j^\dagger f_j(S) ,
 \label{eq.boson2}
\end{eqnarray}
where $a_i$ and $b_j$ are boson annihilation operators, and
\begin{equation}
    f_\ell (S) = \left(1 - \frac{n_\ell}{2S}\right)^{1/2}\\
               = 1 - \frac{1}{2}\frac{n_\ell}{2S} -
                  \frac{1}{8}\left(\frac{n_\ell}{2S}\right)^2 + \cdots ,
\end{equation}
with $n_\ell=a_i^\dagger a_i$ and $b_j^\dagger b_j$.
Indices $i$ and $j$ refer to sites on the \emph{up} and \emph{down} sublattices, 
respectively. Using Eqs.~(\ref{eq.boson1})-(\ref{eq.boson2}),
$H_{\rm mag}$ may be expanded in powers of $1/S$,
\begin{equation}
 H_{\rm mag} = -\frac{1}{2}JS^2Nz + H_{\rm mag}^{(0)} + H_{\rm mag}^{(1)} 
             + H_{\rm mag}^{(2)} + \cdots, 
\end{equation}
where $N$ and $z$ are the number of lattice sites and that of nearest neighbor
sites, respectively. 
The leading term $H_{\rm mag}^{(0)}$ is expressed as
\begin{equation}
 H_{\rm mag}^{(0)} = JS\sum_{<i,j>}( a_i^\dagger a_i + b^\dagger_{j} b_{j}
           +  a_i b_{j} + a_i^\dagger b_{j}^\dagger).
\label{eq.h0}
\end{equation}   
The Fourier transforms of the boson operators are introduced 
within the first magnetic Brillouin zone (MBZ),
\begin{eqnarray}
 a_i &=& \left(\frac{2}{N}\right)^{1/2}\sum_{\textbf{k}} a_{\textbf{k}}
  \exp(i \textbf{k}\cdot \textbf{r}_i), \\
 b_j &=& \left(\frac{2}{N}\right)^{1/2}
       \sum_{\textbf{k}} b_{\textbf{k}} \exp(i \textbf{k}\cdot \textbf{r}_j).
\end{eqnarray}
Then, we introduce a Bogoliubov transformation,
\begin{equation}
 a_{\textbf{k}}^\dagger = \ell_{\textbf{k}}\alpha_{\textbf{k}}^\dagger
          + m_{\textbf{k}}\beta_{-\textbf{k}}, \quad
 b_{-\textbf{k}} = m_{\textbf{k}}\alpha_{\textbf{k}}^\dagger
          + \ell_{\textbf{k}}\beta_{-\textbf{k}}, 
\label{eq.magnon}
\end{equation}
with
\begin{eqnarray}
 \ell_{\textbf{k}} &=& \Bigl[\frac{1+\epsilon_{\textbf{k}}}
 {2\epsilon_{\textbf{k}}}\Bigr]^{1/2},\quad
  m_{\textbf{k}} = -\Bigl[\frac{1-\epsilon_{\textbf{k}}}
  {2\epsilon_{\textbf{k}}}\Bigr]^{1/2} \equiv - x_{\textbf{k}}\ell_{\textbf{k}},
\label{eq.xk}\\
 \epsilon_{\textbf{k}} &=& \sqrt{1-\gamma_{\textbf{k}}^2}, \quad
 \gamma_{\textbf{k}} = \frac{1}{z}\sum_{\mbox{\boldmath{$\delta$}}}
     {\rm e}^{i{\textbf{k}} \cdot \mbox{\boldmath{$\delta$}} },
\end{eqnarray}
where $\mbox{\boldmath{$\delta$}}$ connects the origin with the nearest 
neighbor sites.
By this transformation, $H_{\rm mag}^{(0)}$ is diagonalized as
\begin{equation}
 H_{\rm mag}^{(0)} = JSz\sum_{\textbf{k}}(\epsilon_{\textbf{k}}-1) 
     + JSz\sum_{\textbf{k}} \epsilon_{\textbf{k}}
   (\alpha_{\textbf{k}}^\dagger \alpha_{\textbf{k}}
   + \beta_{\textbf{k}}^\dagger\beta_{\textbf{k}}).
\end{equation}
Similarly,  $H_{\rm mag}^{(1)}$ is expressed as
\begin{widetext}
\begin{eqnarray}
 H_{\rm mag}^{(1)} &=& \frac{JSz}{2S} A\sum_{\textbf{k}}\epsilon_{\textbf{k}}
 (\alpha_{\textbf{k}}^\dagger \alpha_{\textbf{k}}
+ \beta_{\textbf{k}}^\dagger\beta_{\textbf{k}})
 \nonumber \\
     &+&\frac{-JSz}{2SN}\sum_{1234}\delta_{\textbf{G}}(1+2-3-4)
     \ell_1\ell_2\ell_3\ell_4  
\nonumber \\
     &\times& 
\biggl[\alpha_1^\dagger\alpha_2^\dagger\alpha_3\alpha_4 
     B_{1234}^{(1)}+\beta_{-3}^\dagger\beta_{-4}^\dagger\beta_{-1}\beta_{-2} 
     B_{1234}^{(2)}+4\alpha_1^\dagger\beta_{-4}^\dagger\beta_{-2}\alpha_3 
     B_{1234}^{(3)}  \nonumber \\
     &+&\bigl(2\alpha_1^\dagger\beta_{-2}\alpha_3\alpha_4 B_{1234}^{(4)}
      +2\beta_{-4}^\dagger\beta_{-1}\beta_{-2}\alpha_3 B_{1234}^{(5)}
      +\alpha_1^\dagger\alpha_2^\dagger\beta_{-3}^\dagger\beta_{-4}^\dagger
     B_{1234}^{(6)} + {\rm H.c.}\bigr)\biggr].
\label{eq.intham}
\end{eqnarray}
\end{widetext}
Here the first term in Eq.~(\ref{eq.intham}) is known as the Oguchi correction,
\cite{Oguchi60} which coefficient $A$ is given by
\begin{equation}
 A=\frac{2}{N}\sum_{\textbf{k}}(1-\epsilon_{\textbf{k}}) .
 \label{eq.A}
\end{equation}
For the square lattice, $A=0.1579$.
The second term represents the interaction between magnons, where
$\textbf{k}_1$, $\textbf{k}_2$, $\textbf{k}_3$, $\cdots$ are abbreviated
as $1,2,3,\cdots$, and the Kronecker delta $\delta_{\textbf{G}}(1+2-3-4)$ 
indicates the conservation of momenta within a reciprocal lattice vector 
$\textbf{G}$. The vertex functions $B^{(i)}$'s in a symmetric parametrization
are given in Ref.~\onlinecite{Harris71} and \onlinecite{Igarashi92-1}. 
For example, $B^{(3)}_{1234}$, which describes the scattering of two magnons,
is given by
\begin{eqnarray}
 B^{(3)}_{1234}&=&\gamma_{2-4}+\gamma_{1-3}x_1 x_2 x_3 x_4+\gamma_{1-4}x_1 x_2
               + \gamma_{2-3}x_3 x_4 \nonumber\\
	       &-&\frac{1}{2}(\gamma_2 x_4+\gamma_1 x_1 x_2 x_4
	        +\gamma_{2-3-4}x_3+\gamma_{1-3-4}x_1 x_2 x_3 
	 \nonumber\\
	       &+&\gamma_4 x_2 + \gamma_3 x_2 x_3 x_4 + \gamma_{4-2-1}x_1 
	          +\gamma_{3-2-1}x_1 x_3 x_4) .\nonumber \\
 \label{eq.A16}
\end{eqnarray}

The second-order term $H_{\rm mag}^{(2)}$ is composed of products of six 
boson operators.
Writing it in a normal product form with respect to spin-wave operators,
we have 
\begin{eqnarray}
 H_{\rm mag}^{(2)} &=& \frac{JSz}{(2S)^2}\sum_{\bf k}\left[ C_1({\bf k})
  (\alpha_{\bf k}^\dagger \alpha_{\bf k} 
+ \beta_{\bf k}^\dagger\beta_{\bf k}) \right. \nonumber \\
& & \left.
       + C_2({\bf k})(\alpha_{\bf k}^\dagger\beta_{\bf -k}^\dagger
             + \beta_{\bf -k}\alpha_{\bf k})+\cdots \right].
\end{eqnarray}
Neglected terms are unnecessary for calculating corrections up to the
second order. The explicit forms of $C_1({\bf k})$ and $C_2({\bf k})$,
are given by Eqs.~(2.22) and (2.23) in Ref.~\onlinecite{Igarashi92-1}.
       
\subsubsection{the Green function}
We introduce the Green functions for spin-waves,
\begin{eqnarray}
 G_{\alpha\alpha}({\bf k},t) &=& -i \langle T(\alpha_{\bf k}(t)
 \alpha_{\bf k}^\dagger (0)) \rangle, \\
 G_{\alpha\beta}({\bf k},t) &=& -i \langle T(\alpha_{\bf k}(t)
 \beta_{-{\bf k}}(0)) \rangle, \\
 G_{\beta\alpha}({\bf k},t) &=& -i \langle T(\beta_{-{\bf k}}^\dagger(t)
 \alpha_{\bf k}^\dagger (0)) \rangle, \\
 G_{\beta\beta}({\bf k},t) &=& -i \langle T(\beta_{-{\bf k}}^\dagger(t)
 \beta_{-{\bf k}}(0)) \rangle,
\end{eqnarray}
where $ \langle \cdots \rangle$ denotes the expectation
value over the ground state, and 
{\textit T} is the time-ordering operator.
Measuring energies in units of $JSz$, the unperturbed propagators 
corresponding to $H_{\rm mag}^{(0)}$ are given by
\begin{eqnarray}
 G_{\alpha\alpha}^0({\bf k},\omega) &=& [\omega 
 - \epsilon_{\bf k} + i\delta]^{-1}, \\
 G_{\alpha\beta}^0({\bf k},\omega) &=& G_{\beta\alpha}^0({\bf k},\omega) = 0, 
 \\
 G_{\beta\beta}^0({\bf k},\omega) &=& [-\omega 
 - \epsilon_{\bf k} + i\delta]^{-1}.
\end{eqnarray}
The self-energy is defined by a matrix 
Dyson's equation\cite{Harris71}
\begin{eqnarray}
 G_{\mu\nu}({\bf k},\omega) &=& G_{\mu\nu}^0({\bf k},\omega)
  \\
&+& \sum_{\mu'\nu'} G_{\mu\mu'}^0({\bf k},\omega)
 \Sigma_{\mu'\nu'}({\bf k},\omega)G_{\nu'\nu}({\bf k},\omega). \nonumber
\end{eqnarray}
It is expanded in powers of $1/(2S)$,
\begin{equation}
 \Sigma_{\mu\nu}({\bf k},\omega) = \frac{1}{2S}
    \Sigma_{\mu\nu}^{(1)}({\bf k},\omega)
 + \frac{1}{(2S)^2}\Sigma_{\mu\nu}^{(2)}({\bf k},\omega) + \cdots. 
\end{equation}
The first-order terms are obtained from $H_1$:
\begin{equation}
\begin{array}{lcl}
 \Sigma_{\alpha\alpha}^{(1)}({\bf k},\omega)
  = \Sigma_{\beta\beta}^{(1)}({\bf k},\omega) &=& A\epsilon_{\bf k},\\
 \Sigma_{\alpha\beta}^{(1)}({\bf k},\omega)
   = \Sigma_{\beta\alpha}^{(1)}({\bf k},\omega)&=& 0.\\
\end{array}
\end{equation}
The second-order term $\Sigma_{\mu\nu}^{(2)}({\bf k},\omega)$ is given by
the second-order perturbation:
\begin{widetext}
\begin{eqnarray}
 \Sigma_{\alpha\alpha}^{(2)}({\bf k},\omega)
  &=& C_1({\bf k}) + \left(\frac{2}{N}\right)^2
  \sum_{{\bf p}{\bf q}}2\ell_{\bf k}^2\ell_{\bf p}^2
  \ell_{\bf q}^2\ell_{{\bf k}+{\bf p}-{\bf q}}^2  
\left[\frac{\mid B_{{\bf k},{\bf p},{\bf q},[{\bf k+p-q}]}^{(4)}
 \mid^2}{\omega-\epsilon_{\bf p}-\epsilon_{\bf q}-\epsilon_{\bf k+p-q}+i\delta}
 -\frac{\mid B_{{\bf k},{\bf p},{\bf q},[{\bf k+p-q}]}^{(6)}\mid^2}
{\omega+\epsilon_{\bf p}+\epsilon_{\bf q}+\epsilon_{\bf k+p-q}-i\delta}\right]
\nonumber \\
  &=& \Sigma_{\beta\beta}^{(2)}(-{\bf k},-\omega), 
\label{eq.self1}\\
 \Sigma_{\alpha\beta}^{(2)}({\bf k},\omega)
  &=& C_2({\bf k}) + \left(\frac{2}{N}\right)^2
  \sum_{{\bf p}{\bf q}}2
  \ell_{\bf k}^2\ell_{\bf p}^2\ell_{\bf q}^2\ell_{{\bf k}+
 {\bf p}-{\bf q}}^2 {\rm sgn}(\gamma_{\bf G}) 
B_{{\bf k},{\bf p},{\bf q},[{\bf k+p-q}]}^{(4)}
          B_{{\bf k},{\bf p},{\bf q},[{\bf k+p-q}]}^{(6)}
 \frac{2(\epsilon_{\bf p}+\epsilon_{\bf q}+\epsilon_{{\bf k}+{\bf p}-{\bf q}})}
 {\omega^2-(\epsilon_{\bf p}+\epsilon_{\bf q}+\epsilon_{\bf k+p-q})^2+i\delta}
\nonumber \\
 &=& \Sigma_{\beta\alpha}^{(2)}(-{\bf k},-\omega),
\label{eq.self2} 
\end{eqnarray}
where $\delta\to 0$, and $[{\bf k+p-q}]$ stands for ${\bf k+p-q}$ reduced to
the 1st MBZ by a reciprocal vector ${\bf G}$, that is, 
${\bf [k+p-p']}={\bf k+p-p'}-{\bf G}$. In deriving Eqs.~(\ref{eq.self1})
and (\ref{eq.self2}), we have used the relations,
\begin{equation}
\begin{array}{l}
B_{[{\bf k+p-q}],{\bf q},{\bf p},{\bf k}}^{(5)} =
{\rm sgn}(\gamma_{\bf G})B_{{\bf k},{\bf p},{\bf q},[{\bf k+p-q}]}^{(4)},\\
B_{{\bf q},[{\bf k+p-q}],{\bf k},{\bf p}}^{(6)} =
{\rm sgn}(\gamma_{\bf G})B_{{\bf k},{\bf p},{\bf q},[{\bf k+p-q}]}^{(6)}.  \\
\end{array}
\end{equation}
\end{widetext}

\bibliographystyle{apsrev}
\bibliography{paper}

\begin{thebibliography}{47}
\bibitem[{Com()}]{Com3}
\bibinfo{note}{In the derivation, the Born approximation has been used to the
  core-hole potential, which is not weak. The approximation has partly been
  justified by examining higher-order corrections. See
  Ref.~\onlinecite{Igarashi06}.}
\end{thebibliography}

\end{document}